%% file: arxiv.tex
\documentclass[twocolumn,astrosymb]{aastex631}
\usepackage{booktabs}
\usepackage{ragged2e}
\usepackage{appendix}
\usepackage[figuresright]{rotating}

\newcommand{\drc}[1]{\texttt{drc}}
\newcommand{\kms}[1]{km s$^{-1}$}

\begin{document}
\title{Deep Hubble Space Telescope Photometry of LMC and Milky Way Ultra-Faint Dwarfs: \\A careful look into the magnitude-size relation}

\author[0000-0002-3188-2718]{Hannah Richstein}
\affiliation{Department of Astronomy, University of Virginia, 530 McCormick Road,
Charlottesville, VA 22904 USA}

\author[0000-0002-3204-1742]{Nitya Kallivayalil}
\affiliation{Department of Astronomy, University of Virginia, 530 McCormick Road,
Charlottesville, VA 22904 USA}

\author[0000-0002-4733-4994]{Joshua D.~Simon}
\affil{Observatories of the Carnegie Institution for Science, 813 Santa Barbara Street, Pasadena, CA 91101, USA} 

\author[0000-0001-9061-1697]{Christopher T.~Garling}
\affiliation{Department of Astronomy, University of Virginia, 530 McCormick Road,
Charlottesville, VA 22904 USA}
\author[0000-0003-0603-8942]{Andrew Wetzel}
\affiliation{Department of Physics and Astronomy, University of California, One Shields Avenue, Davis, CA 95616, USA}
\author[0000-0003-1634-4644]{Jack T.~Warfield}
\affiliation{Department of Astronomy, University of Virginia, 530 McCormick Road,
Charlottesville, VA 22904 USA}
\author[0000-0001-7827-7825]{Roeland P.~van der Marel}
\affiliation{Space Telescope Science Institute, 3700 San Martin Drive, Baltimore, MD 21218, USA}
\affiliation{The William H. Miller III Department of Physics \& Astronomy, Bloomberg Center for Physics and Astronomy, \\Johns Hopkins University, 3400 N. Charles Street, Baltimore, MD 21218, USA}
\author[0000-0001-6529-9777]{Myoungwon Jeon}
\affiliation{School of Space Research, Kyung Hee University, 1732 Deogyeong-daero, \\Yongin-si, Gyeonggi-do 17104, Republic of Korea}
\affiliation{Department of Astronomy \& Space Science, Kyung Hee University, 1732 Deogyeong-daero, \\ Yongin-si, Gyeonggi-do 17104, Republic of Korea}
\author[0000-0002-2628-0237]{Jonah C.~Rose}
\affiliation{Department of Astronomy, University of Florida, 211 Bryant Space Science Center, Gainesville, FL 32611, USA}
\affil{Center for Computational Astrophysics,  Flatiron Institute, 162 Fifth Avenue, New York, NY 10010, USA}
\author[0000-0002-5653-0786]{Paul Torrey}
\affiliation{Department of Astronomy, University of Virginia, 530 McCormick Road,
Charlottesville, VA 22904 USA}

\author[0009-0006-2183-9560]{Anna Claire Engelhardt}
\affil{George Mason University, Department of Physics and Astronomy, 4400 University Dr., \\MSN: 3F3, Fairfax, VA 22030, USA}

\author[0000-0003-0715-2173]{Gurtina Besla}
\affiliation{Steward Observatory, University of Arizona, 933 N. Cherry Ave, Tucson, AZ 85719, USA}

\author[0000-0003-1680-1884]{Yumi Choi}
\affil{NOIRLab, 950 N. Cherry Ave, Tucson, AZ 85719, USA}

\author[0000-0002-7007-9725]{Marla~Geha}
\affiliation{Department of Astronomy, Yale University, P.O. Box 208101, New Haven, CT 06520, USA}
\author[0000-0001-8867-4234]{Puragra Guhathakurta}
\affiliation{Department of Astronomy and Astrophysics, University of California Santa Cruz, \\University of California Observatories, 1156 High Street, Santa Cruz, CA 95064, USA}

\author[0000-0001-6196-5162]{Evan N. Kirby}
\affil{Department of Astronomy, California Institute of Technology, 1200 E. California Blvd., MC 249-17, \\Pasadena, CA 91125, USA}

\author[0000-0002-9820-1219]{Ekta~Patel}
\affiliation{Department of Physics and Astronomy, University of Utah, 115 South 1400 East, \\Salt Lake City, Utah 84112, USA}

\author[0000-0001-5618-0109]{Elena Sacchi}
\affil{Leibniz-Institut für Astrophysik Potsdam, An der Sternwarte 16, 14482 Potsdam, Germany}
\affil{INAF--Osservatorio di Astrofisica e Scienza dello Spazio di Bologna, Via Gobetti 93/3, I-40129 Bologna, Italy} 

\author[0000-0001-8368-0221]{Sangmo Tony Sohn}
\affiliation{Space Telescope Science Institute, 3700 San Martin Drive, Baltimore, MD 21218, USA}

\begin{abstract}
We present deep Hubble Space Telescope (HST) photometry of ten targets from Treasury Program GO-14734, including six confirmed ultra-faint dwarf galaxies (UFDs), three UFD candidates, and one 
likely globular cluster. Six of these targets are satellites of, or have interacted with, the Large Magellanic Cloud (LMC). We determine their structural parameters using a maximum-likelihood technique. 
Using our newly derived half-light radius ($r_h$) and $V$-band magnitude ($M_V$) values in addition to literature values for other UFDs, we find that UFDs associated with the LMC do not show any systematic differences from Milky Way UFDs in the luminosity-size plane.
Additionally, we convert simulated UFD properties from the literature into the $M_V-r_h$ observational space to examine the abilities of current dark matter (DM) and baryonic simulations to reproduce observed UFDs. Some of these simulations adopt alternative dark matter models, thus allowing us to also explore whether the $M_V-r_h$ plane could be used to constrain the nature of DM. 
We find no differences in the magnitude-size plane between UFDs simulated with cold, warm, and self-interacting dark matter, but note that the sample of UFDs simulated with alternative DM models is quite limited at present.
As more deep, wide-field survey data become available, we will have further opportunities to discover and characterize these ultra-faint stellar systems and the greater low surface-brightness universe.
\end{abstract}


\section{Introduction} \label{sec:intro}

Within the cosmological paradigm of Cold Dark Matter (CDM), structures in the Universe that we observe today were formed through hierarchical evolution \citep[e.g.,][]{white1978,peebles1984,frenk1988}. The most massive dark matter (DM) halos grew by accreting less massive objects, a phenomenon that has left evidence in many forms, such as galaxy clusters and the satellite population of our own Milky Way (MW). 
While the two most massive satellites of the MW, the Large and Small Magellanic Clouds (LMC, SMC), have been known for millennia, the advent of wide-field optical imaging surveys with digital CCD detectors has enabled the discovery of dwarf satellites orders of magnitude fainter than the Magellanic Clouds.

Among these galaxies are ultra-faint dwarfs (UFDs), which were first discovered after the Sloan Digital Sky Survey (SDSS) began \citep[e.g.,][]{willman2005}, and over 60 candidates have been discovered since \citep[e.g.,][]{belokurov2006,bechtol2015,drlica2015,kim2015,koposov2015a,laevens2015a,homma2016,torrealba2016}. While UFDs seem to be an extension of classical dwarf galaxies ($M_{\star} \approx 10^{5-7} M_{\odot}$), as a sub-class they have been defined as having $L\leq10^5 L_{\odot}$ and $M_{\star} \approx 10^{2-5} M_{\odot}$ \citep[e.g.,][]{bullock2017,simon2019}. 
UFDs have the oldest, least chemically enriched stellar populations and could be the ``fossils'' of the first galaxies, having formed at least $\sim$80\% of their stellar mass by the end of reionization \citep[$z\sim6$;][]{madau2014} with little-to-no star-formation activity since \citep[e.g.,][]{bovill2009,brown2014,weisz2014a,sacchi2021}. 

The infall history of these UFDs is encoded in their present-day dynamics. While some information can be extracted with only line-of-sight velocities \citep[e.g.,][]{Rocha2012, Carlin2019, Garling2020}, full 6-D kinematics like those provided by the Gaia mission \citep{gaia1} allow for full orbital reconstruction. This in turn allows us to search for signatures of group infall.
With Gaia proper motions (PMs), a subset of UFDs have been linked kinematically with the LMC \citep[e.g.,][]{kallivayalil2018, patel2020,erkal2020}, and could thus be considered ``satellites of satellites.'' These UFDs have occupied a different environment than the non-LMC associated dwarf galaxies for much of their existence. 

For example, the LMC-associated UFDs had been in proximity to a moderately massive ($\gtrsim 10^{11} M_{\odot}$) galaxy before their infall into the MW halo. Additionally, with the LMC being on its first infall \citep[e.g.,][]{besla2007,kallivayalil2013}, these satellites may have been farther away from the proto-MW during the epoch of reionization ($z\sim6$) than the MW satellites that were captured prior to the LMC's infall. Both sets (the non-LMC-associated and LMC-associated) of UFDs would have been distant enough to be at least outside the virial radius of the MW \citep{wetzel2015,santistevan2023}.

While essentially all UFDs appear to have formed the majority of their present-day stellar mass by the end of reionization \citep[e.g.,][]{brown2014,weisz2014a}, UFDs that inhabited different environments during reionization could have different quenching times due to the inhomogeneity of reionization \citep{Dawoodbhoy2018,Ocvirk2020,Sorce2022,kim2023}. As such, comparing the star formation histories (SFHs) of UFDs that inhabited different environments during reionization could offer insights into this epoch.
Splitting the UFDs into subgroups based on their LMC association, \cite{sacchi2021} found tantalizing signs of differences in the SFHs. 
The LMC-associated UFDs continued forming stars for, on average, more than 500 Myr after the non-LMC-associated galaxies had completed their star formation. This could hint at the galaxies' early environment having an impact on their evolution.

There are some properties of galaxies that we might expect to be largely invariant to environment and more dependent on their DM halo mass. For example, the existence of a stellar-to-halo-mass relation has been well-established \citep[e.g.,][]{behroozi2010}. However, for halo masses below $\sim10^{10} M_{\odot}$ the relation has been more difficult to characterize \citep[e.g.,][]{Brook2014,GK2014,GK2017,Read2017, munshi2021}. 
It is not immediately clear then, whether equal mass halos in an LMC versus MW environment would (without any disruptive events) evolve to host galaxies of similar stellar mass. Stellar mass is inferred from a galaxy's luminosity, but with such a small sample size, it could be difficult to interpret how meaningful the differences between populations are. 

The goal of this study is thus two-fold: 1) to compare LMC and MW UFDs in an observed space, namely the absolute magnitude ($M_V$) versus half-light radius ($r_h$) plane, and 2) to explore whether simulated UFDs can offer insights about what we observe and what we might expect to observe as newer facilities come online. While reviewing the state of simulated UFDs, we also compare simulation results from baryonic runs based on alternative dark matter models, particularly as they relate to predictions of cored or cuspy density profiles. 

The ``core-cusp'' problem arose from the mismatch of the predicted ``cuspy'' density profiles from cold DM-only simulations and the implied ``cored'' profiles from observed rotation curves \citep[e.g.,][]{flores1994,moore1994}, and has been one of the classical challenges to the CDM model \citep[see e.g.,][for a review]{bullock2017,sales2022}. While solutions have since been presented that are consistent with CDM \citep[e.g.,][]{navarro1996, gelato1999, read2005, oh2011, pontzen2012, chan2015, lazar2020, orkney2021}, we take this opportunity to examine and compare the predictions coming from alternative DM models on UFD mass scales.

Here, we choose to revisit the seven targets from \cite{sacchi2021}--Horologium~I (Hor~I), Hydra~II (Hya~II), Phoenix~II (Phe~II), Reticulum~II (Ret~II), Sagittarius~II (Sgr~II), Triangulum~II (Tri~II), and Tucana~II (Tuc~II)--as well as add three additional targets from the Hubble Space Telescope (HST) Treasury Program GO-14734 (PI: N. Kallivayalil) that could be associated with the LMC: Grus~II (Gru~II), Horologium~II (Hor~II), and Tucana~IV (Tuc~IV) \citep{battaglia2022,correaMagnus2022}. We note that Hor~II has not been spectroscopically confirmed to be a dark-matter dominated galaxy and that Sgr~II is most likely a globular cluster \citep{longeard2021,Baumgardt2022}. We include Sgr~II to remain consistent with the original \cite{sacchi2021} sample and will refer to all of the included targets as UFDs for ease of discussion.

The structure of this paper is as follows.
In Section \ref{sec:phot}, we describe the observations and photometry process. In Sections \ref{sec:struct} and \ref{sec:results}, we discuss our methods of structural analysis and absolute magnitude calculations, then examine our results in the context of past literature. Section \ref{sec:disc} includes our comparison of LMC-associated and non-LMC associated UFDs, as well as a comparison of observed and simulated UFDs. We conclude in Section \ref{sec:conc}.

\input{obsTable}

\section{Photometry of Hubble Space Telescope Data} \label{sec:phot}
\subsection{Target Selection}
As explained above, we choose to study the seven UFDs whose SFHs were characterized in \cite{sacchi2021}. In this work the seven UFDs were separated into LMC (Hor I, Phe II, Ret II) and non-LMC associated UFDs (Hya II, Sgr II, Tri II, Tuc II) on the basis of orbit models \citep{patel2020} constructed with Gaia DR2 proper motions for the UFDs \citep{kallivayalil2018, erkal2020}.
Previous studies \citep[e.g.,][]{jethwa2016,sales2017} also supported some of these classifications, but where there were disagreements, the results of \cite{patel2020} were adopted as they used the most updated PM information (Gaia DR2) available at the time. 

Since then, \cite{battaglia2022}, \cite{correaMagnus2022}, and \cite{pace2022}, among others, have published studies using Gaia eDR3 \citep{GaiaEDR3} to improve the PM measurements and orbital models for several of these UFDs. \cite{battaglia2022} support the classification of Hor~I, Phe~II, and Ret~II as ``highly likely former satellites of the LMC,'' and adds Hor~II and Tuc~IV as ``potential" former satellites and Gru~II as a recently captured LMC satellite. \cite{correaMagnus2022} also present orbital models which suggested Tuc~IV and Gru~II were recently captured by the LMC, while Hor~II and Ret~II only show a moderate likelihood of having been associated with the LMC in the past.
Other suspected/confirmed LMC satellites include Carina II, Carina III, and Hydrus I \citep{kallivayalil2018,erkal2020,patel2020,battaglia2022,correaMagnus2022}, however we do not present photometry for them here as they were not observed as part of the Treasury program.

\subsection{Observations and Reduction}
All ten targets (Table \ref{t:obs}) were observed in the F606W and F814W filters of the HST Advanced Camera for Surveys \citep[ACS;][]{ford1998} using the Wide Field Channel (WFC). 
Off-target fields were simultaneously observed with the UV/Visible channel (UVIS) of the Wide Field Camera 3 (WFC3) in order to characterize the stellar background/foreground distributions. Each ACS/WFC field-of-view (FOV) is $202\arcsec \times 202\arcsec$, while the WFC3/UVIS off-fields are $162\arcsec \times 162\arcsec$. 
As these observations were planned as first epochs for follow-up PM measurements, each target field had four dithered exposures in order to facilitate well-sampled templates for background galaxies. 
These were taken using two orbits in each filter, with an integration time per exposure of approximately one half-orbit ($\sim$1100 s). Table \ref{t:obs} gives the average exposure time in integer seconds for each target. 
Hor~II, Tri~II, and Tuc~IV had two pointings, while Tuc~II had four pointings, so as to ensure there were enough stars for robust PM measurements. 

The individual dither (\texttt{flc}) images were downloaded from the Mikulski Archive for Space Telescopes, having already been processed and corrected for charge transfer inefficiency using the current ACS and WFC3 pipelines \citep{Anderson2018,Anderson2021}. The four separate dithers from each filter were aligned to each other using TweakReg and then combined using the \texttt{DRIZZLE} algorithm \citep{Fructer2002} as implemented by the HST software \textsc{Drizzle}Pac \citep{stsci2012,avila2015} to create \drc{} fits files. The \texttt{segmentation} routine from \texttt{photutils} was used to identify and mask the chip gaps, detector artifacts, and large saturated sources.

\subsection{DAOPHOT-II}

The photometry will be described in detail in the data paper accompanying the public release of the PSF-fit source catalogs (H. Richstein et al., 2024, in prep.). We provide a brief overview here.

PSF-fitting photometry was performed on the masked \drc{} files using \texttt{DAOPHOT-II} and \texttt{ALLSTAR} \citep{stetson1987,stetson1992}. The source lists from F606W and F814W were matched using \texttt{DAOMATCH} and \texttt{DAOMASTER} to create a preliminary PSF source catalog. To calibrate our photometry, we first removed \texttt{DAOPHOT}'s intrinsic zeropoint of 25 mag, then added appropriate VegaMag zeropoints (from the ACS Zeropoint Calculator for ACS and \texttt{stsynphot} for WFC3) corrected for the exposure times of our images.                    

The PSF magnitudes are computed within a finite radius, so to correct for the infinite aperture case, we applied an aperture correction.
To calculate this, we performed aperture photometry on the same \drc{} images using the \texttt{photutils} \citep{Bradley2020} routines \texttt{DAOStarFinder} and \texttt{aperture\_photometry}. We used an aperture with a radius of 0.2\arcsec and applied the VegaMag zeropoints and encircled energy corrections from \cite{bohlin2016} and \cite{medina2022} for ACS and WFC3/UVIS, respectively. The F606W and F814W aperture source lists were then matched to each other to create an aperture photometry catalog. Sources were matched between the PSF and aperture photometry catalogs to determine an aperture correction that was then applied to the \texttt{DAOPHOT} PSF-derived VegaMag magnitudes.

\subsection{Artificial Star Tests and Flag Creation}
To determine our photometric completeness limits, photometric error functions, and to aid in developing metrics for star-galaxy separation,
we performed artificial star tests (ASTs) with the \texttt{ADDSTAR} function of \texttt{DAOPHOT}. For each filter and pointing, we inject 500 artificial stars into the science image at a time in order to avoid changing the total source density of the frame significantly. After injecting the artificial stars, we performed the same \texttt{ALLSTAR} photometry as on the masked \drc{} files. We performed this process 500 times so that we had a total of 250,000 artificial stars per filter and pointing.

A primary source of contamination in deep photometric studies of MW dwarf satellites are faint, unresolved background galaxies. As such, essentially all downstream analyses (e.g., measuring structural parameters) rely on methods to exclude these contaminants in order to obtain a highly pure sample of stars in the dwarf galaxy of interest. 
We created a flag to distinguish between stars and galaxies based on the standard error (ERR), SHARP, and CHI parameters of stars recovered from the ASTs. 
The \texttt{DAOPHOT} ERR value is an estimate of the star's standard error, which takes into account the flux, PSF residuals, and noise properties of the images. SHARP and CHI are quality diagnostics relating how the model PSF compares to each source.

The SHARP value of an isolated star should be close to zero, while more extended objects would have SHARP values much greater than zero, and image defects and cosmic rays would have values much less than zero.
The CHI value represents the ratio between the pixel-to-pixel scatter in the PSF residuals and the scatter expected from the image noise properties. Stars should scatter around unity, although brighter stars may have higher values due to saturation effects.

We used exponential functions to divide between acceptable and poor parameter values.
To begin, we binned the artificial star parameters in magnitude space, then calculated the median values across the bins. To account for the spread in parameter space of the observed stars, we added the $3\sigma$ of each bin to its respective median, and then fit an exponential function to those values. For the SHARP parameter, we fit two functions, as there was a spread in both the positive and negative direction. For the negative function, we subtracted instead of added the $3\sigma$.

We assigned sources their flag values (between 0 and 1, inclusive) based on where their SHARP, ERR, and CHI values fell in relation to the functional fits. The ERR and CHI functions acted as upper limits, while the source's SHARP value had to fall between the positive and negative function lines to count as meeting the criteria. If a source met the SHARP and ERR criteria in both F606W and F814W, it was assigned a 1. If sources did not meet these criteria in both filters, but did meet at least the SHARP and ERR criteria in the same filter and the CHI criterion in either filter, they were assigned flag values between 0.5 and 1 (exclusive). For this analysis, we only considered sources with flag values greater than 0.5.

\begin{figure*}
    \centering
    \includegraphics[width=0.8\textwidth]{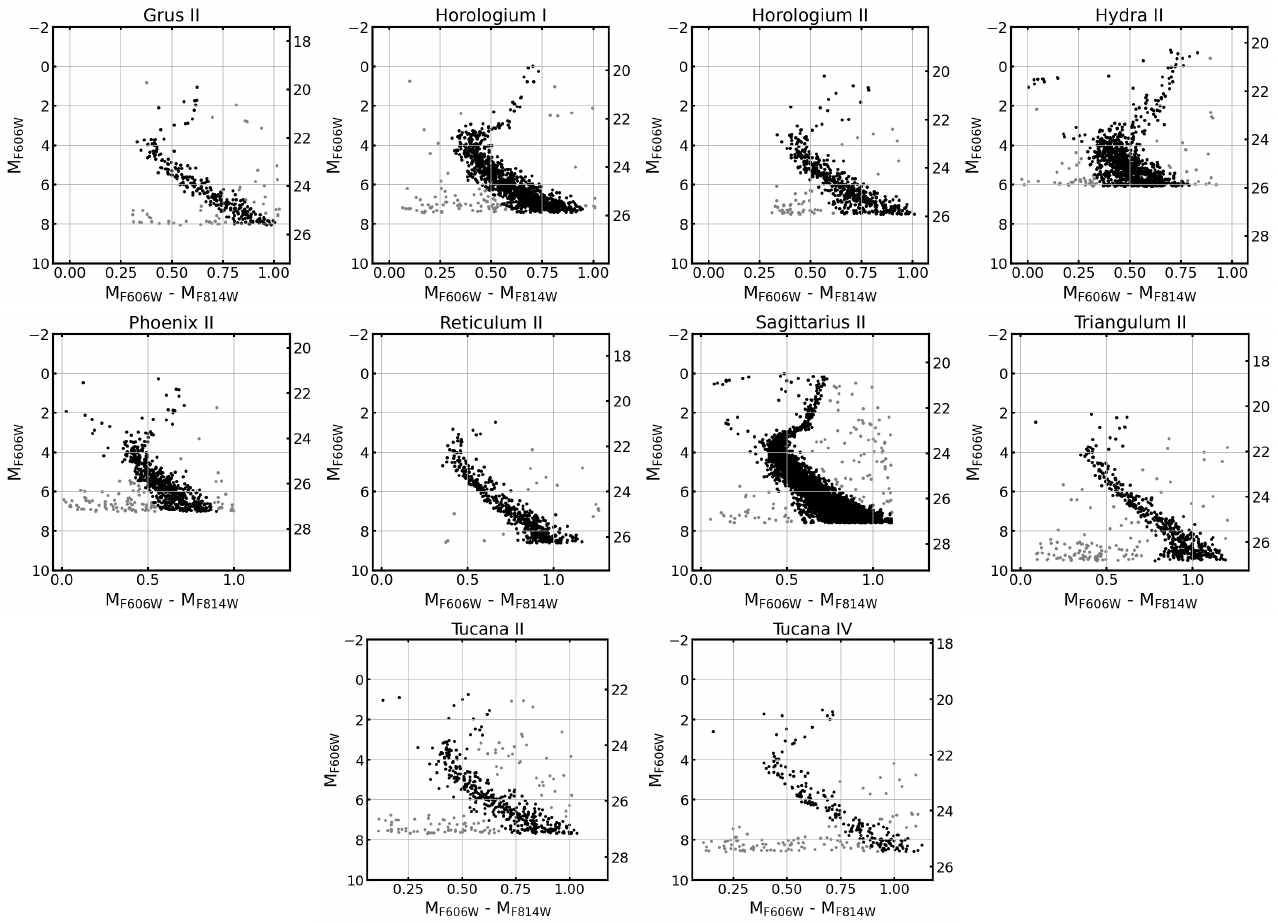}
    \caption{Color-magnitude diagrams in the VegaMag system for the ten targets presented herein. The sources presented are all sources that passed the quality, completeness, and color cuts imposed for the morphological analysis. We have colored in gray the sources that are most likely background/foreground contamination but are used in the structural fitting to determine the background parameter.}
    \label{fig:cmd}
\end{figure*}

\subsection{Color-Magnitude Diagrams}
The color-magnitude diagrams (CMDs) for each target are shown in Figure \ref{fig:cmd}. Here, they have been transposed to absolute magnitude space by applying the dust correction and subtracting the distance modulus listed in Table \ref{tab:derived}. To correct for the reddening and dust extinction, we used the \cite{Schlegel1998} maps within the \texttt{dustmaps} module \citep{Green2018} and applied the \cite{Schlafly2011} recalibration. The correction was calculated for each individual star based on its right ascension (RA) and declination (DEC) values. The sources plotted in the CMDs are those above the 90\% completeness limit and with flag values above 0.5. 
We additionally applied a straight (vertical) color-cut in CMD space tailored to each target for background galaxy removal, and have highlighted the likely member stars in black. We leave the foreground/background sources that pass our color-cut, as the code that we use for fitting structural parameters can account for a foreground/background term. The lack of bright red-giant branch stars in some of the CMDs (e.g., Ret~II, Tri~II) is due to saturation in the images.

\section{Structural Analysis} \label{sec:struct}
In the years following the discovery of the first UFDs, it became important to quantify how well one could actually describe the morphology of such diffuse objects.
In a comprehensive exploration of systematics that affect the measurement of dwarf morphologies with resolved stars, \cite{munoz2012} found three aspects of observational design that influence the measurement accuracy. They found that morphologies are best measured with observations covering a large field of view (FOV; $>$3 times the galaxy half-light radius) to sufficient depth to achieve a central density contrast (relative to the background and foreground contamination) of $\sim$20 and to measure as many stars in the dwarf as possible (preferably $>$1000). 

In practice, ground-based observatories with large fields-of-view can achieve sufficient depths to robustly measure the morphologies of the brighter and closer MW satellites. However, for the faintest and farthest UFDs, these observatories can be limited by the photometric depth that they can achieve and their ability to achieve good enough seeing for confident star-galaxy separation.
Although its FOV is limited, HST/ACS can thus have an advantage in this regime for its superior photometric depth and image resolution, which allows for better star-galaxy separation and lower background levels.

With the available HST pointings, we are able to measure structural parameters for six of the ten targets: Hor~I, Hor~II, Hya~II, Phe~II, Sgr~II, and Tri~II.
For Gru~II, Ret~II, Tuc~II, and Tuc~IV, our imaging covers only between $\sim0.5-1\times$ the half-light radii of these objects, preventing us from making reliable morphological measurements for these objects.

\begin{figure*}
     \centering
    \includegraphics[width=0.75\textwidth]{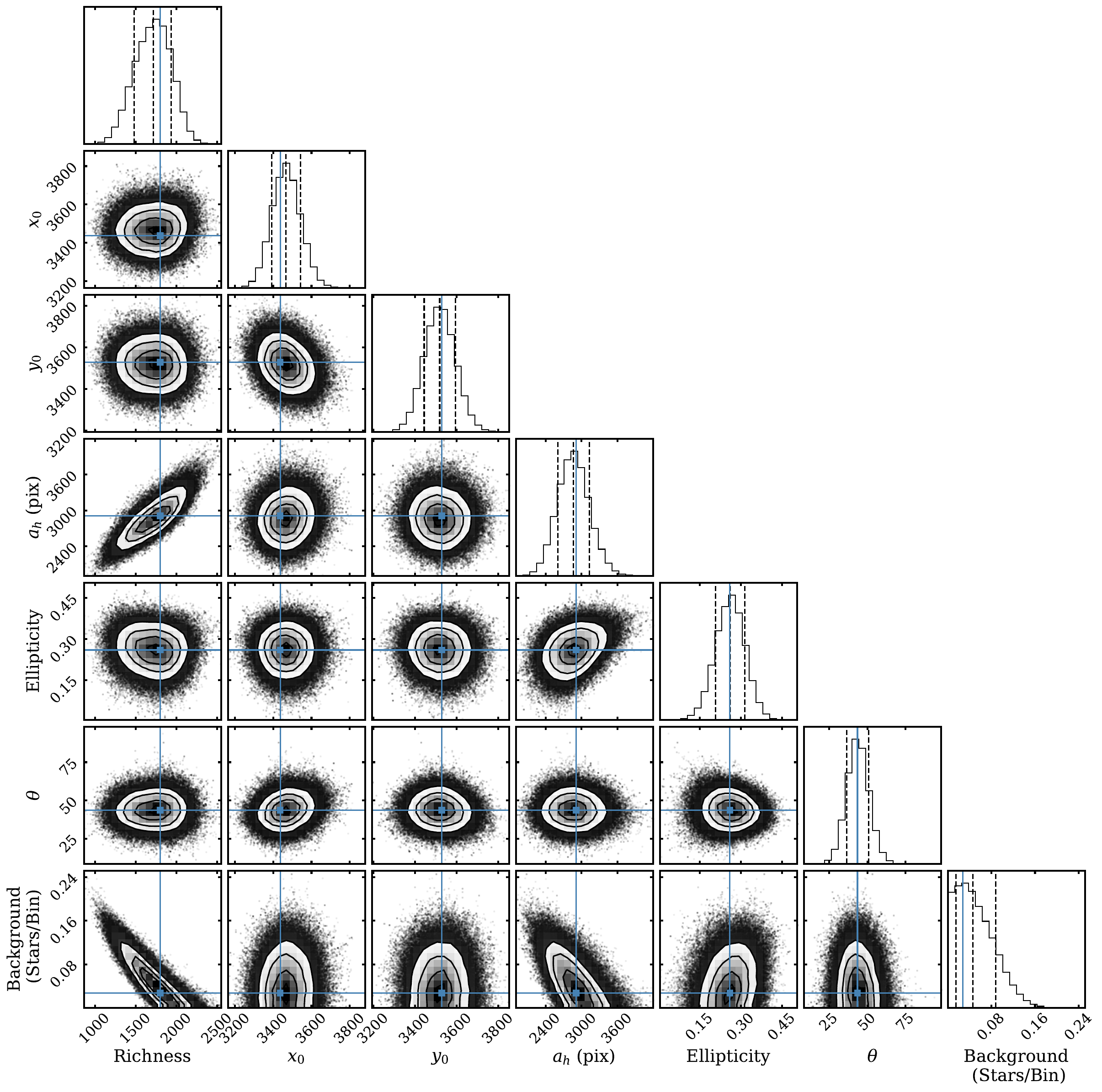}
    \caption{The posterior distributions of the 7-parameter structural fit for Hor~I using the exponential model. The three black vertical dashed lines represent the 16th, 50th, and 84th percentiles. The blue lines and markers show the maximum likelihood values.}
    \label{fig:corner}
\end{figure*}

\subsection{Fitting 2D Profiles} \label{sec:fitting}
We employed the same technique used in \cite{DrlicaWagner2020} and \cite{Simon2021} that was based on the maximum likelihood algorithm presented in \cite{martin2008} to fit 2D profiles of each galaxy. For each target, we modeled both an exponential and a Plummer \citep{Plummer1911} profile, then performed binned Poisson maximum likelihood fits to the probability density functions. We had seven free parameters: center position ($x_0,y_0$), number of stars (richness), 2D projected semi-major axis of the ellipse containing one-half the surface density\footnote{Similar to past morphology studies \citep[e.g.,][]{martin2008}, assuming there is no mass segregation, the half-density and half-light radii are equivalent.} of the galaxy (elliptical half-light radius; $a_h$), ellipticity\footnote{Ellipticity is defined as $\epsilon=1-b/a$, where $a$ and $b$ are the lengths of the semi-major and semi-minor axes, respectively.} ($\epsilon$), position angle of the semi-major axis measured from North through East ($\theta$), and foreground/background surface density (average density of star-like sources in the FOV not belonging to the galaxy; $\Sigma_b$). 

As we noted in \cite{richstein2022}, much of the past literature \citep[e.g.,][]{martin2008,munoz2018} reporting structural parameters has used $r_h$ when referring to the elliptical half-light radius, though $r_h$ has also been used to report azimuthally-averaged half-light radius values 
\citep[equal to $a_h\sqrt{1-\epsilon}$; e.g.,][]{drlica2015}.
Thus, we choose to make the distinction here, and in our tables of past measurements use $a_h$ for measured elliptical half-light radii. We use $r_h$ as the azimuthally-averaged half-light radius in our later comparison to simulations.

We divided the ACS FOV into $4.55\arcsec \times 4.55\arcsec$ bins and applied masks to account for the ACS chip gap, saturated foreground stars, and extended, resolved background galaxies and to match the mask applied during the photometry process. We counted the stars in each bin and used the Markov chain Monte Carlo (MCMC) ensemble sampler \texttt{emcee} \citep{ForemanMackey2013} to sample the posterior probability distribution.
We refer the reader to \cite{richstein2022} for the explicit functional forms\footnote{We note that the in-text equation for $r_e$ in \cite{richstein2022} (\S3.1) should be $r_e=a_h/1.68$.} used for the exponential and Plummer profiles. 
Figure \ref{fig:corner} shows an example corner plot \citep{corner} with the posterior distributions for the exponential fit to Hor~I. The corner plots for the exponential fits of the other targets are shown in Appendix \ref{app:corner}.

We see three different correlations between pairs of parameters: richness and background, $a_h$ and background, and richness and $a_h$.
These relations come from the finite number of stars used in the fit. For example, richness and background counts both depend on whether individual stars are considered to be contributing the surface density of the galaxy or not. If more stars belong to the galaxy, the richness will increase and the background will be lower. Similarly, a higher richness correlates with a larger $a_h$, as a greater number of stars at extended distances from the center would be considered part of the galaxy. This $a_h$ and richness correlation also explains the $a_h$ and background anti-correlation, as a larger $a_h$ with a higher richness necessitates fewer stars belonging to the background.

As shown in \cite{wheeler2019}, measuring dwarf morphologies from shallow imaging with surface brightness limits $<30$ mag arcsec$^{-2}$ can bias size measurements towards being too compact. For typical dwarfs in our sample, we are able to trace their light profiles to $\sim31.2$ mag arcsec$^{-2}$ such that we should experience no significant bias in our morphological measurements from our surface brightness limits. It is more common in our measured sample to be limited by our imaging FOV, as our imaging for some objects does not meet the recommended coverage of three times the $a_h$ \citep{munoz2012}.

\subsection{Apparent and Absolute V-band Magnitude Calculations}
To calculate integrated magnitudes, we transformed the extinction-corrected F606W and F814W VegaMag values to $V$ and $I$ band, respectively. For the ACS photometry, we used the \cite{Sirianni2005} conversions and for the WFC3 values, we used the coefficients from \cite{harris2018}.

We applied the same probabilistic model used in \cite{richstein2022}. Briefly, this entailed creating a Gaussian kernel (for both the ACS and WFC3 fields) with \texttt{scikitlearn} \citep{scikit-learn} \texttt{Kernel Density} and fitting it to the stars in CMD space. Random samples were generated from these kernels and the log-likelihood of each sample was computed with respect to the binned CMD data. We calculated excess flux from non-member sources using a probabilistic background model created from the WFC3 field. 
The stellar density was determined by multiplying the log-likelihoods by the area of the CMD box and subtracting the off-field (WFC3) model from the on-field (ACS) model. The integral returned the flux of the stars inside the CMD box, which we multiplied by a correction factor to account for the flux outside the FOV. We converted this flux into magnitude space, giving the integrated apparent magnitude ($m_V$).

We performed this calculation within a Monte Carlo simulation to include magnitude errors of the individual sources and the uncertainties on the FOV correction incorporating different model parameters. The median integrated $m_V$ values are presented in Table \ref{tab:derived}. To calculate the absolute integrated magnitude ($M_V$), we ran the same Monte Carlo simulation but included the distance modulus and its associated uncertainty. 

As our methodology for measuring the integrated magnitudes of our galaxies does not account for the luminosity from stars fainter than our observational completeness limits, we are likely underestimating the total luminosity of our targets. Using a stellar luminosity function constructed by combining the Kroupa (\citeyear{kroupa2001}) initial mass function with a 12 Gyr, [Fe/H]$=-2.2$ PARSEC \citep{bressan2012,chen2014,marigo2017} isochrone, we estimate that for the targets considered here, we are missing between $1$--$10\%$ of their total luminosities with an average of $\sim3\%$. This indicates that our integrated magnitudes may be biased fainter by at most $\sim0.1$ mag due to our photometric completeness limits. The other sources of uncertainty (FOV correction, distance) thus dominate the reported uncertainties.

\subsection{Mass-to-Light Ratios}
We used our measured elliptical half-light radii to calculate updated mass estimates with velocity dispersion values ($\sigma_v$) coming from the literature (see Table \ref{tab:derived}). We employed the \cite{Wolf2010} equation:
\begin{equation}
    M_{1/2} \simeq \frac{4}{G} \sigma^2_v R_e.
\end{equation}
The $R_e$ in this equation is defined as the 2D-projected half-light radius from elliptical fits, which is equivalent to our $a_h$.
$M_{1/2}$ is the mass enclosed within this $R_e$.

The $V$-band luminosity was calculated by converting the integrated $V$-band magnitude using a Solar $M_V=4.83$. We divided the $M_{1/2}$ calculated above by the resulting $L_V$ for the mass-to-light ratio. Uncertainties reported in Table \ref{tab:derived} are the 84th and 16th percentiles from the Monte Carlo simulation using the individual uncertainties on the mass and luminosity.

\begin{figure}
    \centering
    \includegraphics[width=0.45\textwidth]{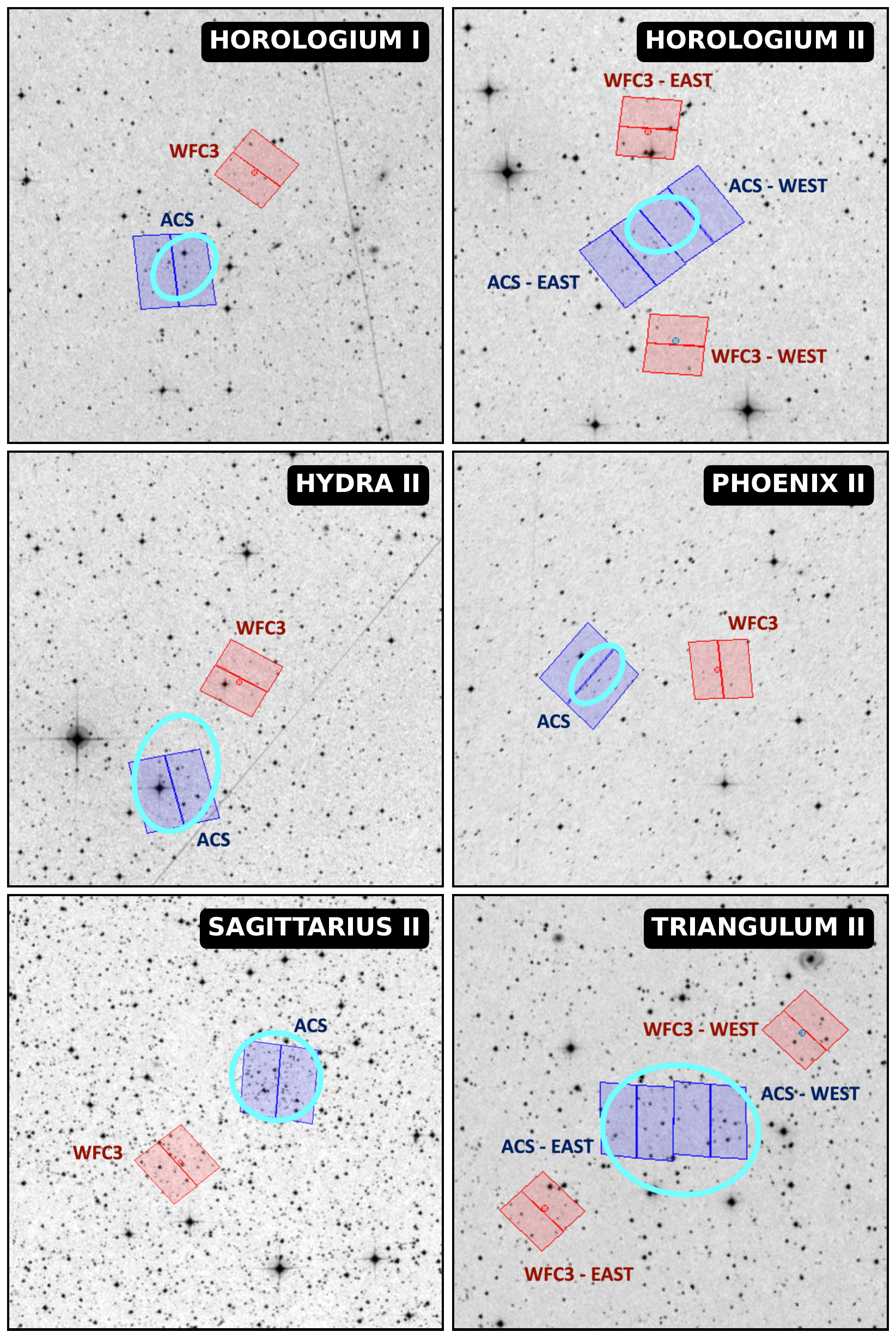}
    \caption{Digitized Sky Survey cutout images from Aladin showing the relative size of our measured half-light ellipses ($a_h$; semi-major axis) and HST fields-of-view. North is up and East is to the left for all panels.
    The ACS FOV in blue has a side length of 202\arcsec; the WFC3 FOV in red has a side length of 162\arcsec. For targets with multiple ACS pointings, the corresponding WFC3 off-field is labeled to match the ACS label, not the WFC3 relative physical positions. While our FOV/$a_h$ ratios are less than the \cite{munoz2012} recommendation ($\sim$3), we are generally able to recover $a_h$ values within 2$\sigma$ agreement of past literature measurements.}
    \label{fig:fov1}
\end{figure}

\input{derivedTable}

\begin{figure*}
     \centering
    \includegraphics[width=0.8\textwidth]{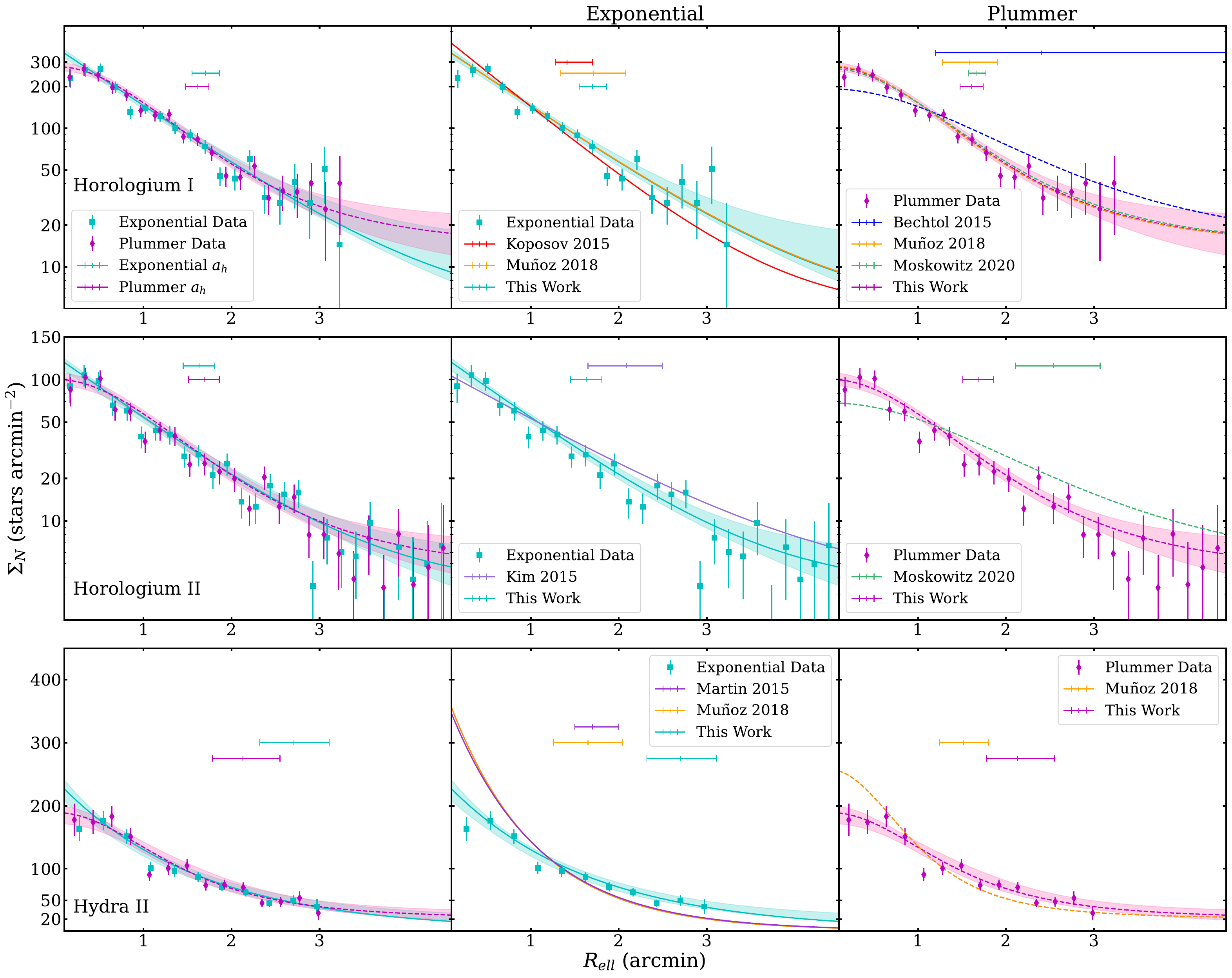}
    \caption{The 1D surface density profiles of Hor~I (top row), Hor~II (middle), and Hya~II (bottom) plotted against the elliptical radius, $R_{ell}$. The left column shows the surface density measurements of the annularly-binned data, at increments of 0.1$a_h$ for the respective models (exponential, cyan; Plummer, magenta). Uncertainties are derived from Poisson statistics.
    The plotted curves represent the 1D profiles produced with the best-fit parameters of the 2D data. Floating horizontal lines represent the $a_h$ measurements and their uncertainties. The middle column shows our exponential fits and 1D surface density measurements compared to existing literature measurements. The 1D profiles from literature fits are normalized to have an area under the curve equal to ours. The right column is the same as the middle, except for the Plummer model rather than the exponential. Note the agreement, despite the surface density profiles being fit to 2D-binned data rather than the 1D-binned data also shown here.}
    \label{fig:prof}
\end{figure*}

\begin{figure*}
     \centering
    \includegraphics[width=0.8\textwidth]{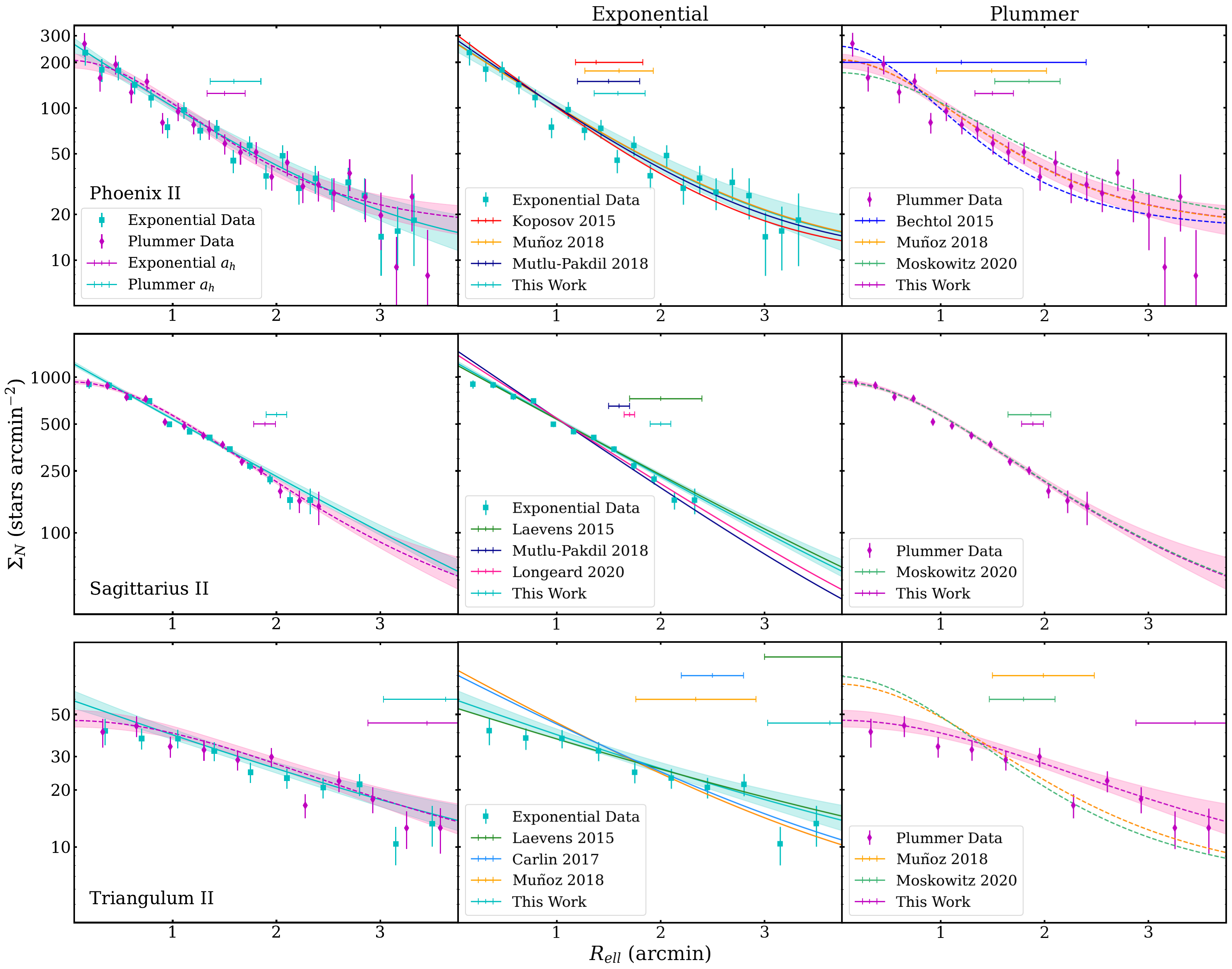}
    \caption{The same as Figure \ref{fig:prof}, for the targets Phe~II (top row), Sgr~II (middle), and Tri~II (bottom).}
    \label{fig:prof2}
\end{figure*}

\section{Results} \label{sec:results}
\subsection{Comparison of Measured Systems}
Here, we report and compare our results to the literature for Hor~I, Hor~II, Hya~II, Phe~II, Sgr~II, and Tri~II. 
Figure \ref{fig:fov1} shows the relative sizes of our measured half-light radii against the observed fields-of-view. We have overplotted a cyan ellipse encompassing 50\% of the surface density of each target (according to our best-fit exponential model) on a Digitized Sky Survey \citep[DSS;][]{dss} optical cutout of the region of interest from Aladin \citep{aladin}. Our ACS footprints are shown in blue, with the WFC3 off-fields in red.

We also discuss their classification status (e.g., star cluster or ultra-faint galaxy) and their possible association with the LMC.
UFD candidates are typically confirmed to be galaxies based on velocity dispersion measurements determined spectroscopically \citep[e.g.,][]{simon2007}, as higher values suggest large mass-to-light ratios. 
The presence of a metallicity dispersion has also been used to support classification as a galaxy, as it suggests more than one period of star formation \citep[][]{willman2012}.
Particularly relevant here is the work of \cite{fu2023}, which estimated photometric metallicity distribution functions (MDF) and dispersion values for a few of the objects in our sample using narrow-band HST imaging. Their analysis showed good agreement with smaller samples of spectroscopic measurements, and we include a discussion of their results for the applicable targets here.

Another classification technique that has recently been applied to UFDs relies on stellar mass segregation \citep{Baumgardt2022}. Energy equipartition should cause high-mass stars to become more centrally concentrated within a cluster, while the low-mass stars move further out. 
The timescale for this process is longer in stellar systems that inhabit dark matter halos. As such, systems which show no sign of stellar mass segregation are more likely to be galaxies than star clusters.
We report the \cite{Baumgardt2022} classification alongside or in lieu of the spectroscopic determination for completeness.

For brevity, we note here that four of our targets (Hor~I and Phe~II (discussed here), Ret~II and Tuc~II (discussed in Section \ref{sec:non-meas})) were discovered concurrently by \cite{koposov2015b} and \cite{bechtol2015} using data from the first year of the Dark Energy Survey (DES). \cite{koposov2015b} reported azimuthally-averaged radii ($r_h$) rather than elliptical semi-major axis lengths ($a_h$) for their physical sizes.\footnote{We note that the $r_{maj}$ values given in \cite{koposov2015b} are the exponential scale lengths, $a_h/1.68$} In our written comparisons, we present these $r_h$ values but also include the physical $a_h$ that we calculated using their $\epsilon$ values. The uncertainties reported on $a_h$ in Tables \ref{tab:hor1}, \ref{tab:phe2}, \ref{tab:ret2}, and \ref{tab:tuc2} are from a Monte Carlo using the \cite{koposov2015b} $r_h$, $\epsilon$, and ($m-M$) values. 

For all of the targets in our paper except for Hya~II, \cite{moskowitz2020} reported ``circularized'' projected half-light radii\footnote{\cite{moskowitz2020} fit both a 1- and a 3-component Plummer; we choose to report the best-fitting values from the 1-component model as that model is most similar to the Plummer function that we adopted.} (our $r_h$) in arcminutes.
To ease comparison, we converted the \cite{moskowitz2020} fit $r_h$ to an $a_h$ in arcminutes using their reported $\epsilon$.
We denote $a_h$ values derived using the exponential model as $a_{h,\mathrm{exp}}$ and those from the Plummer model as $a_{h,\mathrm{p}}$.

Figures \ref{fig:prof} and \ref{fig:prof2} show the 1D surface density model profiles for the six UFDs that we measured, as well as the binned surface density values. Note that these 1D profiles were not fit to the binned data, but rather the result of flattening the best-fit 2D model profiles. The x-axis is the elliptical radius, $R_{ell}$, which can be described as the radial distance when the major axis is aligned with the position angle (see \cite{richstein2022} Eq. 3 for the full mathematical expression).
The left column shows only the measurements from this work, while the middle and right column compare our findings to past literature values for the exponential and Plummer models, respectively. We normalized the areas-under-the-curve from the literature to our area values. 

\subsubsection{Horologium~I}
Both \cite{koposov2015b} and \cite{bechtol2015} were unable to constrain a position angle for Hor~I, and \cite{koposov2015b} were only able to give an upper limit on its ellipticity using the discovery data. The two teams also reported different radii: \cite{koposov2015b} estimated Hor~I to have a $\sim1\farcm3$  azimuthally-averaged radius ($a_{h,\mathrm{exp}}\sim1\farcm4$), while \cite{bechtol2015} reported a Plummer radius of $\sim2\farcm4$ with large uncertainties. The next set of published structural parameters for Hor~I was from \cite{munoz2018}, who fit both Plummer and exponential models to photometry that they performed on the public DES data. \cite{munoz2018} measured an exponential half-light radius of $\sim1\farcm7$, and a Plummer half-light radius of $\sim1\farcm6$. The $a_{h,\mathrm{p}}$ of \cite{moskowitz2020} was $\sim1\farcm7$. 

Our $a_{h,\mathrm{exp}}=1\farcm70^{+0.16}_{-0.15}$ and $a_{h,\mathrm{p}}=1\farcm61\pm0.13$ agree within $1\sigma$ with all past literature measurements. 
Using the previous \cite{munoz2018} $a_{h,\mathrm{exp}}=1\farcm71$ as a benchmark, the FOV/$a_h$ ratio would be just short of 2, though Hor~I was one of the targets for which over 1000 sources were used in the structural fitting.
Our $\theta$ and $\epsilon$ values for both the exponential and Plummer fits additionally agree within 1$\sigma$ with the \cite{munoz2018}. Our empirically-derived $M_V=-3.4\pm0.2$ mag is also in agreement with past literature values.

Shortly after publishing the discovery paper, \cite{koposov2015a} reported on follow-up data of Ret~II and Hor~I taken with the Very Large Telescope (VLT) FLAMES-GIRAFFE spectograph. For Hor~I, they confirmed five member stars and found a kinematic dispersion of $4.9^{+2.8}_{-0.9}$ km s$^{-1}$, corresponding to a mass-to-light ratio of $\sim$570 $M_{\odot}/L_{\odot}$. Though \cite{koposov2015a} also found a low metallicity dispersion (0.17 dex), they find that the low [Fe/H]$=-2.76$ and large mass-to-light ratio made it consistent with the other UFD galaxies known at the time. 

A more detailed spectroscopic analysis based on three members stars using VLT FLAMES-UVES and the MIKE spectograph on the Magellan-Clay Telescope was published by \cite{nagasawa2018}. Their team found a similarly low average [Fe/H] $\sim-2.6$. The \cite{Baumgardt2022} mass-segregation measurement was not significant, supporting the classification of Hor~I as a dwarf galaxy.
From HST narrow-band imaging of 27 stars, \cite{fu2023} calculated an average [Fe/H]$=-2.79^{+0.12}_{-0.13}$ dex and a $\sigma_{\rm [Fe/H]}=0.56^{+0.11}_{-0.09}$ dex. Removing a possible higher-metallicity outlier, the $\sigma_{\rm [Fe/H]}$ dropped to $0.41^{+0.11}_{-0.10}$ dex. Either metallicity spread is consistent with Hor~I being a galaxy.

\subsubsection{Horologium~II}
\cite{kim2015} discovered Hor~II in the DES Y1A1 images, measuring an $a_{h,\mathrm{exp}}=2\farcm09^{+0.44}_{-0.41}$ ($r_{\mathrm{h}}$ in their paper). The only other structural parameter measurement is from \cite{moskowitz2020}; using DES DR1, they fit $a_{h,\mathrm{p}}\sim2\farcm5$, which agreed within 1$\sigma$. 
Hor~II is one of our targets that had two telescope pointings, giving us approximately a combined FOV of $3\farcm3 \times 6\farcm7$. The observations were oriented for maximal coverage along the semi-major axis, making our FOV/$a_h$ ratio (with the \cite{kim2015} value as a reference) slightly over 3 along that direction.

Our $a_{h,\mathrm{exp}}=1\farcm63\pm0.18$  measurement is just outside of being within $1\sigma$ agreement with the \cite{kim2015} value. Our $a_{h,\mathrm{p}}=1\farcm69^{+0.18}_{-0.17}$ is within $1.7\sigma$ of the \cite{moskowitz2020} value. Our exponential $\theta$ and $\epsilon$ values agree with the \cite{kim2015} within 1.5 and 1.6$\sigma$, respectively. For the Plummer model, our $\theta$ and $\epsilon$ are within 1 and 1.1$\sigma$ of the \cite{moskowitz2020} values.
Our integrated $M_V=-2.1 \pm0.2$ mag is 0.5 magnitudes fainter than the $M_V=-2.6$ presented by \cite{kim2015}, but still within $1.8\sigma$ agreement.

There has not yet been a spectroscopic study of Hor~II, but \cite{Baumgardt2022} found no evidence for mass segregation and suggested that Hor~II is likely a dwarf galaxy.

\subsubsection{Hydra~II}
\cite{martin2015} discovered Hya~II serendipitously in data from the Survey of the Magellanic Stellar History \citep[SMASH;][]{nidever2017}, which used the Dark Energy Camera (DECam) on the Blanco 4m telescope. They reported an $a_{h,\mathrm{exp}}=1\farcm7$ ($r_h$ in their paper), with the caveat that the chip gaps and a bright foreground star near the center of the galaxy added uncertainty to the measurement. The \cite{munoz2018} measurement from the DECam data gave an $a_{h,\mathrm{exp}}\sim1\farcm7$ and an $a_{h,\mathrm{p}}\sim1\farcm5$.

Our best-fit $a_{h,\mathrm{exp}}$ is $2\farcm70^{+0.41}_{-0.38}$, and our $a_{h,\mathrm{p}}$ is $2\farcm13^{+0.42}_{-0.35}$, which are within 2 and 1.4$\sigma$ of the \cite{munoz2018} values, respectively. 
Hya~II is another target for which we have over 1000 sources, however, the saturated star mentioned in the \cite{martin2015} discovery paper is also near the center of our frame. 
With the \cite{munoz2018} $a_{h,\mathrm{exp}}=1\farcm65$ as a reference, our FOV/$a_h$ is $\sim$2. 
This discrepancy is thus likely due to the limited FOV, the position of the chip gap, and the saturated star. As this field was not optimized with structural measurements as a science goal, we consider the \cite{munoz2018} measurement based on a wider FOV to be more reliable.

From our fit, we measured a center approximately $45\arcsec$ from the original \cite{martin2015} value. To investigate whether this was significant, or an artifact from our smaller FOV combined with the saturated star, we downloaded the SMASH DR2 object catalog \citep{nidever2021} for the appropriate region. We applied a distance cut to only consider sources within $6\arcmin$ of the \cite{martin2015} center, as well as quality cuts for $|\mathrm{SHARP}|\leq0.4$ and $\chi<2$. Additionally, we required $\texttt{prob}\neq0$, where $\texttt{prob}$ is the average Source Extractor \citep{bertin1996} ``stellaricity probability value.'' We ran our fitting code on the SMASH data and recovered a center that was $\sim40\arcsec$ from the \cite{martin2015} value, but within $1.4\sigma$ of our newly-derived HST value. It is possible that the shallower nature of the discovery data led to a different stellar density center than is fit by the two more recent, deeper data sets.
Our $M_V=-4.6^{+0.2}_{-0.3}$ mag is consistent with the previous measurements from \cite{martin2015} and \cite{munoz2018}.

\cite{kirby2015} presented Keck~II/DEIMOS spectroscopy for 13 member stars of Hya~II. They reported 90 and 95\% confidence levels for the upper limits of the velocity dispersion corresponding to upper limits on the mass-to-light ratios of 200 and 315 $M_{\odot}/L_{\odot}$, respectively. \cite{kirby2015} were able to marginally resolve a metallicity dispersion of $\sim$0.40 dex, and suggest that this supports tentative classification as a galaxy. This dispersion paired with the large physical radius more strongly supports the dwarf galaxy nature of Hya~II, although \cite{simon2019} does not list Hya~II among the spectroscopically confirmed UFDs. 
More recently, \cite{fu2023} measured the MDF of Hya~II from HST imaging and reported an average [Fe/H] of $-3.08^{+0.11}_{-0.12}$ dex and a $\sigma_{\rm [Fe/H]}=0.33^{+0.12}_{-0.13}$ dex based on 30 stars. Including a more metal-rich possible outlier, the $\sigma_{\rm [Fe/H]}$ increased to $0.47^{+0.13}_{-0.12}$ dex.

\subsubsection{Phoenix~II}
From the DES data for Phe~II, \cite{koposov2015b} reported an azimuthally-averaged $r_h$ of $1\farcm3$  ($a_{h,\mathrm{exp}}\sim1\farcm7$), and \cite{bechtol2015} published an $a_{h,\mathrm{p}}$ value of $\sim1\farcm2$, although \cite{bechtol2015} were unable to constrain a position angle or ellipticity. Using the same DES data, \cite{munoz2018} measured similar values, with $a_{h,\mathrm{exp}}\sim1\farcm6$ and $a_{h,\mathrm{p}}\sim1\farcm5$. \cite{mutluPakdil2018} took follow-up data with Megacam at the Magellan Clay telescope, reaching $\sim$2-3 magnitudes deeper, and reported an exponential $a_{h,\mathrm{exp}}\sim1\farcm5$. The most recent literature measurement from \cite{moskowitz2020} used DES DR1 and fit $a_{h,\mathrm{p}}\sim1\farcm9$. 

Our $a_{h,\mathrm{exp}}=1\farcm58^{+0.26}_{-0.23}$ is consistent with the three past literature measurements. Our $a_{h,\mathrm{p}}=1\farcm5^{+0.20}_{-0.17}$ is also consistent with the past Plummer profile fits. Broadly, our $\theta$ and $\epsilon$ values are consistent with past measurements, with the exception of \cite{moskowitz2020}.
Using the \cite{munoz2018} $a_{h,\mathrm{exp}}=1\farcm60$, our FOV/$a_h$ is 2.1. 
We calculate $M_V=-2.9^{+0.2}_{-0.1}$ mag, consistent with all past literature measurements. 

VLT/FLAMES spectroscopy for Phe~II was presented in \cite{fritz2019}, who measured a velocity dispersion of $11.0\pm9.4$ km s$^{-1}$ and an intrinsic metallicity spread of 0.33 dex for the five stars that the authors identified as members. Their assessment of Phe~II as a galaxy agreed with the previous suggestion by \cite{mutluPakdil2018}, who based their classification on its ellipticity and position in the size-luminosity plane.
\cite{Baumgardt2022} found no significant mass segregation, supporting the dwarf galaxy classification; they noted, however, that because of its large relaxation time, there is still the possibility that Phe~II could be a star cluster. 
\cite{fu2023} report a mean [Fe/H]$=-2.36^{+0.18}_{-0.16}$ dex and a $\sigma_{\rm [Fe/H]}=0.41^{+0.22}_{-0.17}$ dex from 10 HST stars, and paired with the mass segregation information, supported the classification of Phe~II as a galaxy.

\subsubsection{Sagittarius~II}
\cite{laevens2015b} reported the discovery of this satellite from the Panoramic Survey Telescope and Rapid Response System \citep[Pan-STARRS 1, PS1;][]{chambers2016} survey data, and measured $a_{h,\mathrm{exp}}\sim2\arcmin$ ($r_h$ in their paper). This satellite's nature (cluster versus galaxy) was ambiguous at the time of discovery leading to the dual name of Sagittarius~II/Laevens~5, however, all subsequent papers adopted the Sgr~II naming. \cite{mutluPakdil2018} presented deeper Magellan Megacam imaging with an $a_{h,\mathrm{exp}}$ of $\sim1\farcm6$, which agreed within the uncertainties with that of \cite{laevens2015b}. \cite{longeard2020} took deep observations with the Canada-France-Hawaii Telescope (CFHT) MegaCam and presented $a_{h,\mathrm{exp}}\sim1\farcm7$ ($r_{\mathrm h}$ in their paper). \cite{moskowitz2020} used Pan-STARRS DR1 \citep{chambers2016} and fit a larger $a_{h,\mathrm{p}}\sim1\farcm9$.

Our $a_{h,\mathrm{exp}}=1\farcm94\pm0.08$ is larger than recent measurements, though consistent with \cite{laevens2015a}.
We differ at a level of $\sim2.6\sigma$ with \cite{longeard2020} and \cite{mutluPakdil2018}. 
Based on the \cite{longeard2020} $a_{h,\mathrm{exp}}=1\farcm7$, our FOV/$a_h$ is a little under 2.
Similarly to \cite{mutluPakdil2018} and \cite{longeard2020}, we also find very little ellipticity, with $\epsilon<0.03$ and $<0.09$ at the 95\% confidence level. 
Additionally, our $a_{h,\mathrm{p}}=1\farcm85^{+0.07}_{-0.08}$ is consistent with the measurement of \cite{moskowitz2020}. 

Reported position angle values vary widely, which could be related to the low ellipticity leading to the lack of a clear semi-major axis. Our measured $\theta$ values agree within 0.1 and 0.6$\sigma$ to the \cite{longeard2020} and \cite{moskowitz2020} values, respectively, however our $\epsilon<0.09$ suggest that $\theta$ may not be meaningful here.
Our $M_V=-5.3\pm0.2$ mag is consistent with \cite{laevens2015a} and \cite{longeard2020}, but not with the \cite{mutluPakdil2018} $M_V=-4.2 \pm0.1$ mag.

\cite{simon2019sgr} presented a spectroscopic study of Sgr~II based on Magellan/IMACS data, reporting a velocity dispersion of 1.6 $\pm0.3$ km s$^{-1}$ and a metallicity dispersion $<0.08$ dex at the 95\% confidence limit, suggesting that Sgr~II could be a globular cluster. From Keck~II/DEIMOS data, \cite{longeard2020} measured a velocity dispersion of 2.7$^{+1.3}_{-1.0}$ km s$^{-1}$ and a combined photometric and spectroscopic metallicity dispersion of 0.12$^{+0.03}_{-0.02}$ dex. \cite{longeard2021} used new VLT/FLAMES data combined with the DEIMOS spectra, measuring a smaller velocity dispersion of 1.7 $\pm0.5$ km s$^{-1}$ from 113 stars. From 15 stars, they calculated an unresolved metallicity dispersion less than 0.20 dex at the 95\% confidence limit. \cite{longeard2021} thus concluded that Sgr~II is a globular cluster. 
\cite{Baumgardt2022} additionally found a high level of mass segregation, supporting the star cluster classification.
While \cite{fu2023} do not present their analysis on Sgr~II, they state that their results are consistent with the globular cluster status.

\subsubsection{Triangulum~II}
This was the first discovered ultra-faint satellite from the PS1 data by \cite{laevens2015a}, who then took follow-up imaging with the Large Binocular Cameras at the Large Binocular Telescope (LBT). Unable to determine whether this satellite was a cluster or dwarf galaxy, \cite{laevens2015a} gave it the double name of Laevens~2 and Triangulum~II. The initial $a_{h,\mathrm{exp}}$ value from \cite{laevens2015b} (reported as $r_h$) was $3\farcm9$, although they included the caveat that with the low contrast of the stellar overdensity to the background, deeper data would provide a stronger measurement. 

\cite{carlin2017} presented further follow-up deep imaging from the Subaru Hyper Suprime-Cam that reached $\sim$2 magnitudes deeper than the discovery data and increased the overdensity-to-background contrast. They derived an $a_{h,\mathrm{exp}}$ of $2\farcm5$, consistent within 1.5$\sigma$ to the \cite{laevens2015b} value.
\cite{munoz2018} derived a similar exponential $a_{h,\mathrm{exp}}\sim2\farcm3$ and a slightly smaller $a_{h,\mathrm{p}}\sim2$ using Pan-STARRS data from the discovery team. \cite{moskowitz2020} measured an $a_{h,\mathrm{p}}\sim1\farcm8$ from the Pan-STARRS data. 

Tri~II had two ACS pointings, giving us an FOV of $\sim3\farcm3\times6\farcm7$, and a maximum possible FOV/$a_h$ value of $\sim2.9$ along the semi-major axis.
We measured $a_{h,\mathrm{exp}}=3\farcm50^{+0.59}_{-0.65}$, larger, but within 1.4$\sigma$ agreement with \cite{carlin2017} and \cite{munoz2018}; our value is smaller than the \cite{laevens2015a} measurement, though consistent within $0.4\sigma$. Our $a_{h,\mathrm{p}}=3\farcm25^{+0.51}_{-0.59}$ measurement is larger but within $1.7\sigma$ agreement with the \cite{munoz2018} Plummer measurement and $2.2\sigma$ agreement with the \cite{moskowitz2020} value. 

Our $\theta$ and $\epsilon$ values are consistent with both the \cite{carlin2017} and \cite{moskowitz2020} measurements.
The $\theta$ and $\epsilon$ values from our exponential models are within 1.3 and 1.6$\sigma$ of the \cite{munoz2018}; for the Plummer model, they are within 0.9 and 0.6$\sigma$, respectively.
Our fits are most consistent with the larger $a_h$ and smaller ellipticity presented in the \cite{laevens2015b} discovery paper.
The low number of stars in this galaxy and the slight gap between the two ACS pointings may have affected our fits, and we advise adopting either the \cite{carlin2017} or \cite{munoz2018} values. 
The integrated $M_V=-1.3 \pm0.2$ mag that we measure is in $1\sigma$ agreement with the three past literature determinations.

Numerous spectroscopic studies of Tri~II have been published \citep{kirby2015tri2,martin2016,venn2017,kirby2017,ji2019}, often with results hinting at a dwarf galaxy nature, but with no definitive conclusions. The most recent spectroscopic study by \cite{buttry2022} combined new MMT/Hechtochelle data with the existing Keck~II/DEIMOS set and presented an upper limit on the velocity dispersion of 3.4 km s$^{-1}$ at the 95\% confidence level. Additionally, with new Gaia proper motion information confirming two questionable member stars used in the \cite{kirby2017} metallicity dispersion measurement, \cite{buttry2022} used their combined data to report a value of 0.46$^{+0.37}_{-0.09}$ dex; this large spread lends stronger support to a Tri~II galaxy classification. Additionally, the \cite{Baumgardt2022} study found no mass segregation and supported this conclusion.

\begin{figure}
    \centering
    \includegraphics[width=0.45\textwidth]{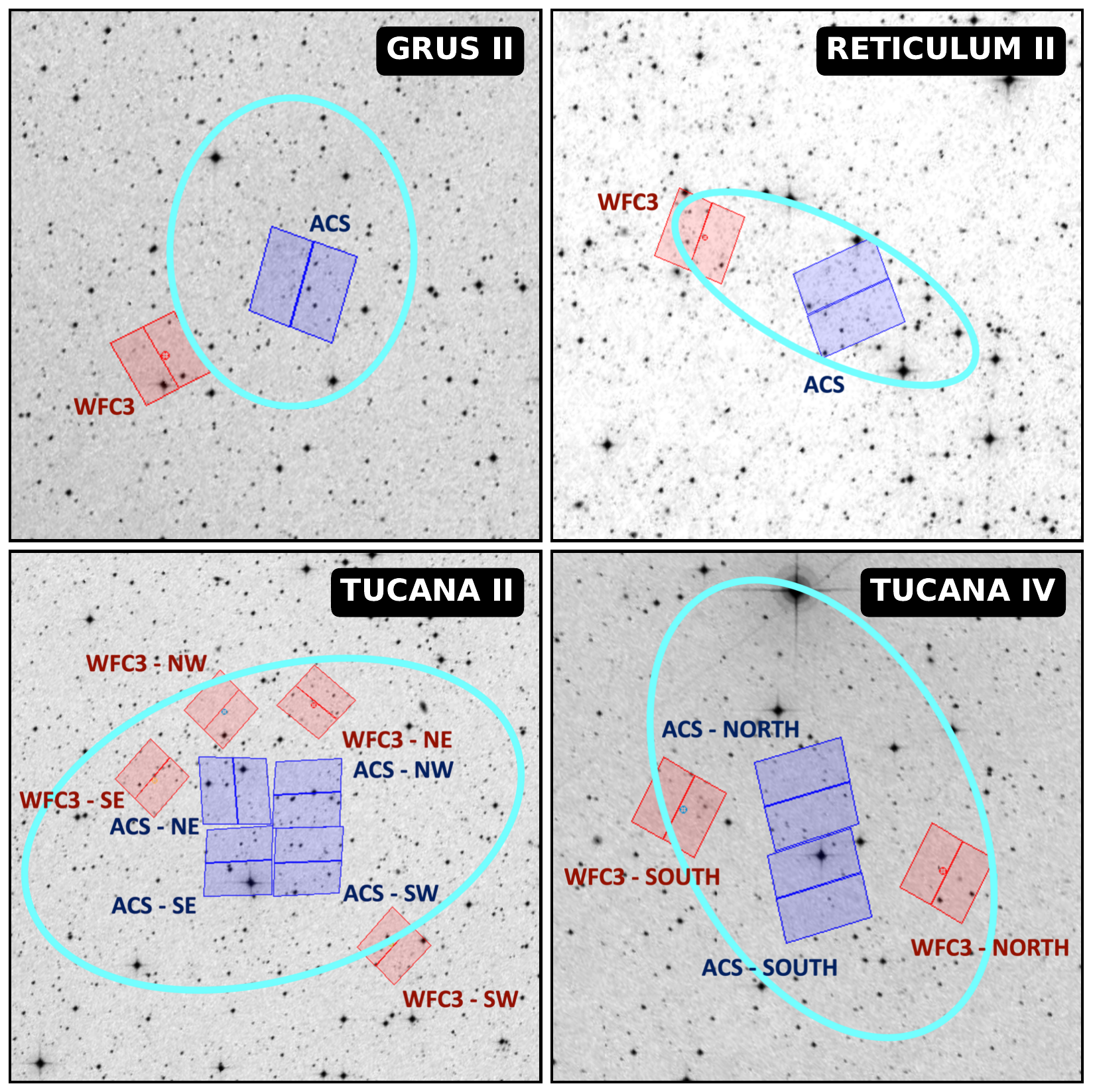}
    \caption{Similar to Figure \ref{fig:fov1}, for the four targets that we present photometry for but do not measure structural parameters. North is up and East is to the left in all panels. The ACS FOV in blue has a side length of 202\arcsec; the WFC3 FOV in red has a side length of 162\arcsec. For targets with multiple ACS pointings, the corresponding WFC3 off-field is labeled to match the ACS label, not the WFC3 relative physical positions.}
    \label{fig:fov2}
\end{figure}

\subsection{Review of Non-Measured Systems} \label{sec:non-meas}

For Gru~II, Ret~II, Tuc~II, and Tuc~IV, we give a brief review of past morphological studies. Figure \ref{fig:fov2} shows the coverage of our HST observations relative to past literature measurements of the target half-light ellipses.

\subsubsection{Grus~II}
Gru~II was the ultra-faint candidate with the highest sigma detection discovered in the second year of DES data, and had no significant ellipticity or asphericity \citep{drlica2015}. As such, there would not be a corresponding $a_h$; \cite{drlica2015} reported $r_h=6\arcmin$. Using the DES Y3A2 catalog,\footnote{DES Y3A2 refers to an internal DES release. The public DES DR1 was based on this \citep{abbott2018}.} \cite{simon2020} repeated the structural analysis and recovered an $a_{h,\mathrm{p}}$ (referred to as $r_{1/2}$) of $5\farcm9$ and an upper limit on the ellipticity of 0.21. 
The \cite{moskowitz2020} $a_{h,\mathrm{p}}$ measurement from DES DR1 \citep{abbott2018} data was larger, at $\sim7\farcm8$, yet within 1.7 and 2.5$\sigma$ of the \cite{drlica2015} and \cite{simon2020} values, respectively.

We did not fit structural parameters for Gru~II, as past measurements of its size have found $a_h\sim6\arcmin$ \citep{drlica2015,moskowitz2020,simon2020}, and our single ACS pointing only extends to $\sim0.25a_h$, which is insufficient for a robust morphology measurement.

The first spectroscopic observation of Gru~II was performed by \cite{simon2020}, who identified 21 member stars with Magellan/IMACS spectroscopy. They were unable to clearly resolve a velocity dispersion, but did report 90 and 95.5\% upper limits. Additionally, \cite{simon2020} did not detect a metallicity spread, however based on the low metallicity ([Fe/H]=$-2.51 \pm0.11$ dex) and large size, they concluded Gru~II is likely a dwarf galaxy. 
Using DOLPHOT photometry of HST data, \cite{Baumgardt2022} did not find significant mass segregation among the member stars, leading them to also conclude that Gru~II is likely a dwarf galaxy.

\subsubsection{Reticulum~II}
From the DES data, in their discovery papers, \cite{koposov2015b} measured an $r_h$ of $\sim3\farcm6$ ($a_{h,\mathrm{exp}}\sim5\farcm7$) and \cite{bechtol2015} measured an $a_{h,\mathrm{p}}\sim6\arcmin$. \cite{munoz2018} fit slightly smaller values of $a_{h,\mathrm{exp}}\sim5\farcm4$ and $a_{h,\mathrm{p}}\sim5\farcm5$ to the same DES data. \cite{mutluPakdil2018} used deeper (by $\sim$2-3 mags) data from Megacam to fit an $a_{h,\mathrm{exp}}\sim6\farcm3$, in agreement with the original \cite{bechtol2015} measurement. The \cite{moskowitz2020} $a_{h,\mathrm{p}}$ from DES DR1 was slightly larger, at $\sim6\farcm6$.
We did not fit structural parameters for Ret~II, as the ACS FOV covers approximately 0.25$a_h$.

Three independent spectroscopic studies of Ret~II were published in the months following the discovery papers \citep{simon2015,walker2015,koposov2015a}, all concluding that Ret~II had the kinematic and spectroscopic properties consistent with being a dwarf galaxy. \cite{simon2015} presented Magellan/M2FS, VLT/GIRAFFE, and Gemini South/GMOS spectroscopy; from 25 member stars, they calculated a velocity dispersion of $3.3 \pm0.7$ km s$^{-1}$ and a metallicity dispersion of $0.28 \pm0.09$ dex. \cite{walker2015} confirmed 17 member stars and reported a velocity dispersion of 3.6$^{+1.0}_{-0.7}$ km s$^{-1}$ and a metallicity dispersion of 0.49$^{+0.19}_{-0.14}$ dex using Magellan/M2FS data. Lastly, \cite{koposov2015a} measured a velocity dispersion of 3.22$^{+1.64}_{-0.49}$ km s$^{-1}$ and a metallicity dispersion of 0.29$^{+0.13}_{-0.05}$ dex from the VLT/GIRAFFE data. 

The most recent spectroscopic study by \cite{ji2023} updated these values to be 2.97$^{+0.43}_{-0.35}$ \kms{} and 0.32$^{+0.10}_{-0.07}$ dex for the velocity and metallicity dispersion, respectively.
\cite{Baumgardt2022} applied their mass-segregation method to Ret~II and found results in agreement with the previous studies, however they noted that due to the large relaxation time, the mass-segregation method alone would not be able to clearly determine the classification.
\cite{fu2023} used HST photometric metallicities from 76 stars to derive an average [Fe/H]$=-2.64^{+0.10}_{-0.11}$ dex and $\sigma_{\rm [Fe/H]}=0.72^{+0.09}_{-0.08}$ dex. This large metallicity dispersion would strongly support a galaxy classification. They note the large discrepancy with past spectroscopic studies and invite follow-up observations for better membership determination.

\subsubsection{Tucana~II}
The discovery parameters for Tuc~II from \cite{koposov2015b} and \cite{bechtol2015} were quite different, with the former reporting an azimuthally-averaged $r_h$ of $9\farcm8$ ($a_{h,\mathrm{exp}}\sim12\farcm9$) and the latter publishing an $a_{h,\mathrm{p}}$ of $7\farcm2$. This discrepancy could be due in part to different CMD masking techniques for member selection. While \cite{bechtol2015} note that there was some elongation in the outer regions of the detected overdensity, they suggest that the distortion was likely noise-related and show that after their CMD-masking, the overdensity is much rounder. Figure 12 of \cite{koposov2015b} shows the CMD-masked stars in the Tuc~II field with the elongation still visible. 
\cite{moskowitz2020} fit DES data for Tuc~II and recovered $a_{h,\mathrm{p}}\sim13\farcm5$ pc, consistent with the $a_{h,\mathrm{exp}}$ of \cite{koposov2015b} within 1$\sigma$.  

Our FOV included four ACS pointings, for a maximum field-width of less than $7\arcmin$. 
We did not fit the structural parameters as though literature $a_h$ measurements varied from $\sim7\arcmin-13\farcm5$, our FOV cover at most the inner one-half of the galaxy.

Using eight probable member stars with spectroscopy from the Michigan/Magellan Fiber System (M2FS), \cite{walker2016} measured a velocity dispersion for Tuc~II of 8.6$^{+4.4}_{-2.7}$ \kms{} and a metallicity dispersion of 0.23$^{+0.18}_{-0.13}$ dex and concluded that it is a dwarf galaxy. The velocity dispersion has since been remeasured by \cite{chiti2023}, who used 16 member stars and calculated a value of 3.8$^{+1.1}_{-0.7}$ \kms{}.

\subsubsection{Tucana~IV}
Tuc~IV had the largest angular size of the ultra-faint satellite candidates discovered in the DES year-two data, with an azimuthally-averaged $r_{\mathrm h}=9.1'$ \citep[$a_{h,\mathrm{p}}\sim11\farcm8$;][]{drlica2015}. The analysis using DES DR1 by \cite{moskowitz2020} measured a slightly smaller $a_{h,\mathrm{p}}\sim9\farcm2$, consistent within 1.3$\sigma$ with the \cite{drlica2015} value. From the DES Y3A2 catalog, \cite{simon2020} published the most recent structural parameters for Tuc~IV, reporting $a_{h,\mathrm{p}}\sim9\farcm3$. Our two ACS pointings would cover approximately the inner one-third of the target, so we did not attempt to characterize the morphology with our HST data.

\cite{simon2020} published the first spectroscopic study of Tuc~IV, significantly resolving a velocity dispersion of 4.3$^{+1.7}_{-1.0}$ \kms{}. From eight RGB stars, they measured a metallicity dispersion of 0.18$^{+0.20}_{-0.18}$ dex. \cite{simon2020} calculated a mass-to-light ratio of 3100$^{+2900}_{-1600}$ $M_{\odot}/L_{\odot}$, indicating a dark-matter dominated nature, and thus spectroscopically confirmed Tuc~IV as a galaxy.

\section{Discussion} \label{sec:disc}
\subsection{Uses for HST Imaging in Structural Analysis}
The HST MW UFD Treasury Program (GO-14734) contains a wealth of data, and precision photometry is essential for optimizing its value. Here, we have presented a subset of the data and an example use case, with a focus on targets that have been kinematically linked to the LMC. 
While a space telescope with a small FOV would not be the premier choice for morphological studies, these types of analyses are still beneficial as ancillary uses for the data.
The depth and resolution of the images have given us the ability to characterize the inner surface densities, and in some cases, recover a full set of structural parameters in agreement with past literature. 
Five out of twelve structural fits (6 targets, 2 models) for $a_h$ fall within 1$\sigma$ of previously measured values, and all fall within $3\sigma$ even though we do not meet the suggested FOV and number of sources criteria \citep[e.g.,][]{munoz2012}.

One of the advantages of high-resolution, space-based imaging is the ability to more clearly distinguish stars from galaxies, as the PSFs are not affected by the atmosphere. This can be important for determining whether apparent structural overdensities from ground-based observations are meaningful or if they are from contaminating background sources. 

An illustrative example of this is when the analysis of ground-based, Megacam imaging of Leo~V seemed to suggest the existence of a debris stream \citep{sand2012} and follow-up space-based observations showed otherwise.
\cite{mutluPakdil2019} presented HST/ACS imaging centered on the overdensity in the Leo~V field, and showed that almost half of the photometric sources were background galaxies. The higher-precision photometry also revealed that many of the true stars from the overdensity were unlikely to be Leo~V members. 

In the case of Hercules, however, photometry of off-center HST/ACS fields contained a main sequence consistent with the stellar population identified in the central pointing, supporting the existence of Hercules' elongation \citep{mutluPakdil2020}. 
The \cite{mutluPakdil2019,mutluPakdil2020} studies both relied on archival imaging of the central regions of their target UFDs, much like the observations we have presented here. 
Their results highlight how we should be cautious in our interpretations of apparent low surface-brightness features, yet at the same time, how important follow-up, space-based imaging is in order to gain more insight into what the structure of a satellite can tell us.

\subsection{The Importance of UFD Structural Characterization}
As UFDs are thought to reside in the least massive DM halos observed to be hosting galaxies \citep[e.g.,][]{bullock2017,simon2019}, their mass accretion histories and their sensitivity to tidal forces have implications for their role in hierarchical structure formation \citep[e.g.,][]{mutluPakdil2020}. 
While we cannot observe directly how UFDs have evolved to their present state, their morphologies may still reflect major events that disturbed their original mass distributions. For example, early morphology studies noted that UFDs were more elliptical than classical dwarfs, with a possible explanation being that these less massive galaxies were more susceptible to tidal effects from the MW \citep[e.g.,][]{martin2008}. While updated measurements from a larger sample size showed that there is little evidence to support a difference between the two populations in this respect \citep{sand2012,simon2019}, the existence of UFD ``tidal features'' (and the need for a physical explanation) remain \citep[e.g.,][]{simon2007,sand2009,munoz2018,mutluPakdil2020}.

One particularly interesting case is that of Tuc~II, one of our limited-FOV targets with only photometry presented. \cite{chiti2021} reported the discovery of member stars out to $\sim$9 times the \cite{bechtol2015} $a_h$ and noted that the orbital parameters of Tuc~II made tidal disruption an incompatible explanation. Soon after, simulation groups were able to reproduce UFD ``tidal features'' without invoking tidal heating from a more massive host galaxy \citep{tarumi2021,goater2023}. For example, assembly from less massive ``building block galaxies'' \citep{tarumi2021} as well as late-time dry accretion \citep{goater2023} can produce extended stellar halos and more elliptical stellar distributions. There has also been more interest in the stellar halos of dwarf galaxies, including how they could be built up through dwarf-dwarf mergers \citep{deason2022} and how to systematically search for them observationally \citep[e.g.,][]{filion2021,jensen2023,longeard2023}.

\begin{figure}
    \centering
    \includegraphics[width=0.99\linewidth]{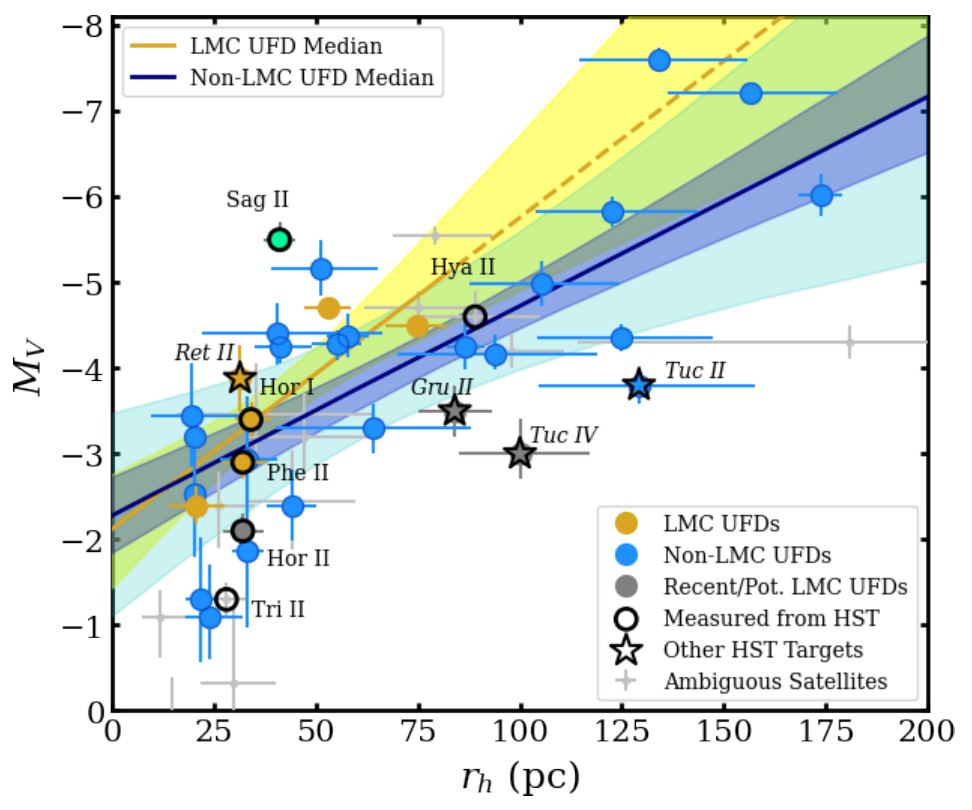}
    \caption{Absolute $V$-band magnitude, $M_V$, versus azimuthally-averaged half-light radius, $r_h$, for UFDs, UFD candidates, and the globular cluster Sag~II (green circle). Golden circles represent the six LMC-associated UFDs, while the blue circles are the non-LMC confirmed UFDs. The three dark gray markers (1 circle, 2 stars) represent the UFDs that are either thought to be recently captured or formerly associated with the LMC, which were not used in the trendline fits.
    Circles with bold black outlines (with labels) show measurements from this work. Star symbols are targets (with italicized labels) for which we present photometry, but do not fit structural parameters. 
    Silver markers with errorbars are non-confirmed UFDs, which were not used in the trendline fits. The solid gold (dark blue) line represents the median value from the Monte Carlo simulations for the LMC (non-LMC) UFD $M_V-r_h$ relation. The dashed continuation of the gold line references the extrapolation of the LMC-associated trendline.
    The gold and dark dark blue shaded regions cover the 16th to 84th percentiles from the Monte Carlos for the respective sets. The cyan shaded region represents the 16th to 84th percentiles of lines fit to the six galaxy, non-LMC Monte Carlo simulation as described in Section \ref{sec:lmcVsMW}.
    The light green band is where the yellow and cyan regions overlap, and the darkest shaded region is where all three uncertainty spaces overlap.
    Though the fit trendlines diverge, the overlapping uncertainty bands suggest no statistical difference.}
    \label{fig:trendlines}
\end{figure}

\subsection{LMC vs. Non-LMC UFDs in the $M_V-r_h$ Plane} \label{sec:lmcVsMW}

\cite{sacchi2021} found tantalizing hints of differences between the SFHs of the LMC- versus non-LMC-associated UFDs, which inspired us to investigate whether the two UFD subgroups showed differences in other observational planes. One could postulate that environmental effects from the density differences in the early Universe or more recently, the MW halo, could have affected the two groups of satellites differently, perhaps leading to either more compact or extended morphologies based on length of time in the MW halo and pericenter distance.

To explore whether there were any differences between the two subgroups apparent in the $M_V-r_h$ parameter space, we fit a line for each set of UFDs to broadly characterize the relationship.
To account for the large uncertainty spaces, we drew $M_V$ and $r_h$ values for each galaxy in the two sets (LMC and non-LMC) using the reported values as the Gaussian mean and the reported uncertainties as the Gaussian sigmas. 
Additionally, we assume an intrinsic scatter in $M_V$ at fixed $r_h$. This is modeled using a Gaussian with a sigma set to the root-mean-square scatter of the data around the best-fit line, with a separately drawn value added to each $M_V$.
For each set of redrawn values, we fit a linear regression model. We performed this procedure 10,000 times per set and show the median $M_V$ value of the evaluated linear functions for $r_h$ from 0 to 200 pc in Figure \ref{fig:trendlines} (gold and dark blue solid lines for the LMC and non-LMC UFDs, respectively). The gold and dark blue shaded regions encompass the 16th to 84th percentiles of $M_V$ values evaluated across the $r_h$ space.

While the median trendlines lines overlap at magnitudes fainter than $-3.5$, at brighter magnitudes they begin to diverge. This seems to agree with the \cite{erkal2020} finding that at fixed luminosity, likely LMC satellites have slightly smaller $r_h$ than MW satellites, which they postulated could be due to different tidal environments.
To test the significance of this comparison, we ran a Monte Carlo that drew 10,000 sets of six random non-LMC associated UFDs with $M_V$ and $r_h$ values drawn using the previously described Gaussian technique. We fit a linear regression to each set of six galaxies and over-plotted the region that encompasses the 16th to 84th percentiles in light blue on Figure \ref{fig:trendlines}. This region overlaps the LMC satellite trendline heavily and encompasses the space around five out of the six LMC UFDs, suggesting that the divergence of the median $M_V-r_h$ relation is not statistically significant. 

\begin{figure*}
    \centering
    \includegraphics[width=0.97\textwidth]{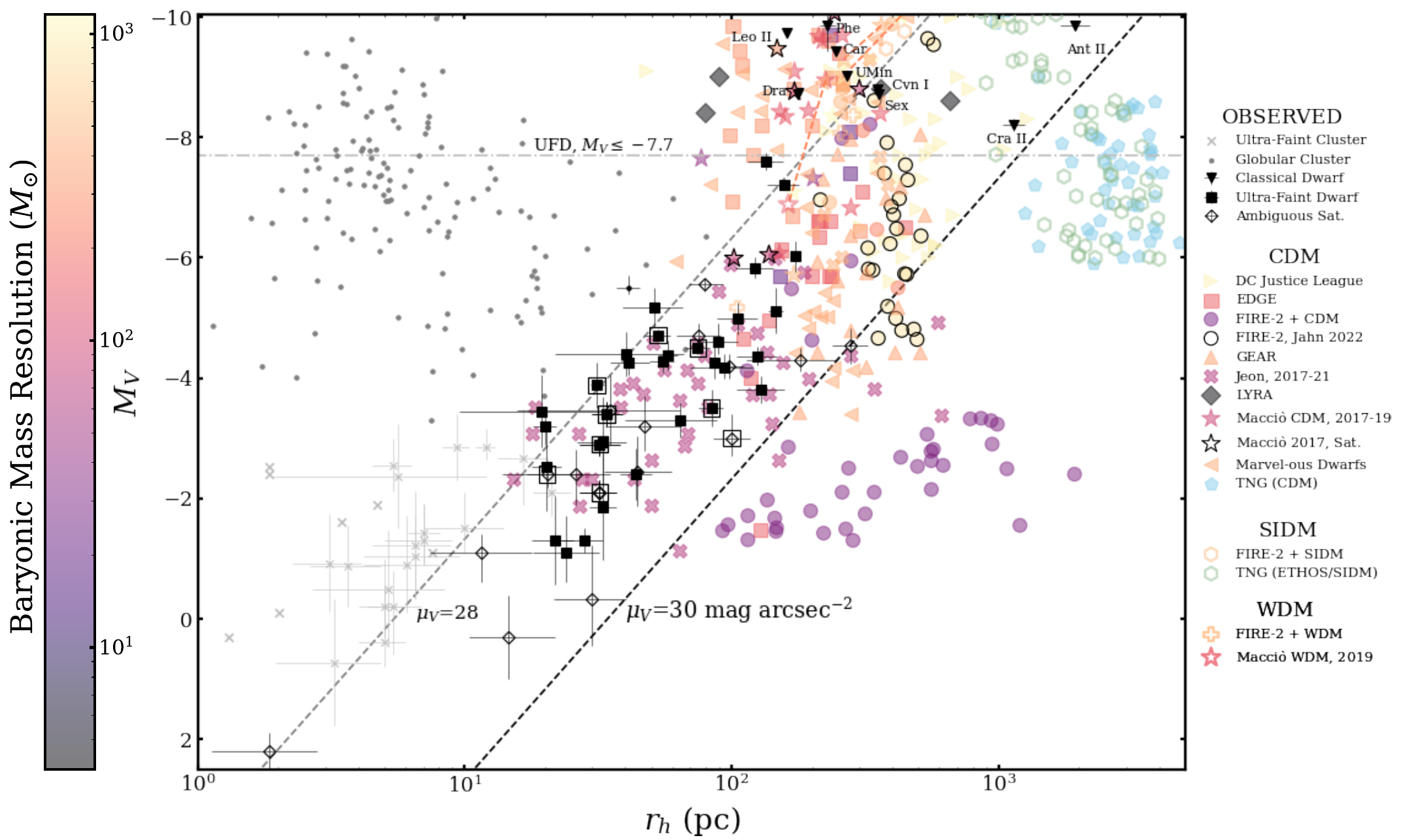}
    \caption{Absolute $V$-band magnitude, $M_V$, versus azimuthally-averaged half-light radius, $r_h$. The dashed light and dark gray diagonal lines represent constant surface brightness values of 28 and 30 mag arcsec$^{-2}$, respectively. The dash-dotted horizontal light gray line marks $M_V\leq-7.7$, often used as the delineation between classical and ultra-faint dwarfs \citep{simon2019}. Classical dwarfs (black inverted triangles) are labelled.
    Measurements of classical dwarfs, UFDs (black squares), and ambiguous satellites (open diamonds) are shown as black markers with error bars. LMC-associated satellites are additionally marked with a black square frame.
    Globular and ``ultra-faint'' clusters are shown as gray points and x's, respectively. All other markers are from simulations and colored by their baryonic particle mass resolution, except for the TNG points, which have a mass resolution $\sim10^4 M_{\odot}$. If different environments were used within the same simulation grouping, we have marked the difference using black outlines.
    Simulated points using alternative DM models are shown as open symbols. The \cite{maccio2019} WDM galaxies are connected to their corresponding CDM galaxies with coral dashed lines.
    Advances in simulations have led many groups to be able to generate galaxy analogs in and around the observed UFD space. Full references for the data are given in Appendix \ref{app:ref}, and references for the simulations are in Table \ref{tab:simtable}. }
     \label{fig:rhMV}
     \label{FIG:RHMV}
\end{figure*}

\input{simTable}

\subsection{Observed and Simulated UFDs in the $M_V-r_h$ Plane}
Baryonic simulations have advanced to the extent that some are now able to resolve galaxy formation in the ultra-faint regime. We present a subset of these simulated UFD-analogs plotted in $M_V$ versus $r_h$ in Figure \ref{fig:rhMV} along with observed UFDs, UFD candidates, and star clusters. 

As simulation groups often report stellar mass ($M_{\star}$) and 3D, half-mass radii rather than $M_V$ and 2D, projected half-light radii, respectively, some literature values had to be converted.
When $M_V$ and 2D $r_h$ values were provided from the simulations, we plotted those.
Otherwise, to place the simulated galaxies in a plane where the quantities are more readily comparable, we converted $M_{\star}$ to $M_V$ using a stellar-mass-to-light ratio of 2, which is reasonable for an ancient stellar population \citep[e.g.,][]{martin2008,simon2019}.
To convert the 3D radius values, we used the relation presented in \cite{Wolf2010}:
\begin{equation}
    R_e \simeq \frac{3}{4} r_{1/2},
\end{equation}

\noindent where $R_e$ is the 2D, projected half-light radius and $r_{1/2}$ is the 3D, deprojected half-light radius. Assuming that all simulated galaxies have roughly spherical mass distributions, $R_e$ is then equivalent to $r_h$.
The simulated galaxies are colored according to the reported baryonic mass resolutions, except for the TNG points, as these were much lower resolution with star particle masses $\sim10^4 M_{\odot}$. Most of the simulations followed the CDM model, but there are some that used warm dark matter (WDM) and self-interacting dark matter (SIDM); these are shown with open markers. Basic properties of the simulations used are listed in Table \ref{tab:simtable}.

For the data, we convert $a_h$ values from the literature to $r_h$ using the reported $\epsilon$ values. Plotted error bars are generated from a Monte Carlo using both the uncertainties on the $a_h$ and $\epsilon$ values. To simplify the plot, we do not show error bars on the globular cluster values from the literature. Star clusters are classified here as ``ultra-faint'' if they have $M_V\geq-3$ \citep[e.g.,][]{cerny2022}. Literature values for classical dwarfs, UFDs, and ambiguous satellites are shown in black. 
For the purposes of this plot, we are not highlighting our new measurements but have included an extra square frame around the LMC-associated sample.

We have also included simulated galaxies from LYRA \citep{gutcke2022} even though no UFDs are produced, as this model created classical dwarf analogs similar to those from the other simulations considered. The LYRA set also has the highest resolution, with a baryonic particle mass of 4 $M_{\odot}$ and is likely going to produce UFD analogs in future studies.

We also note that the TNG galaxies \citep[][J.~Rose et al.~2024, in prep.]{rose2023}~may not be considered direct UFD analogs due to limitations from their use of a pressurized equation of state.
These galaxy formation prescriptions are identical to those used in IllustrisTNG \citep[e.g.,][]{pillepich2018}, a well-known, large volume cosmological simulation that has been shown to be consistent with observed size relations for galaxies with $M_{\star}\gtrsim10^9\ M_{\odot}$ and to have good numerical convergence \citep{genel2018}.
While the TNG galaxies have much lower resolution than the other simulations, 
we include them because they offer a comparison between samples from both CDM and alternative DM models with the same baryonic physics. The ETHOS alternative DM models \citep{vogelsberger2016} included contain both suppression to the initial matter power spectrum in the form of dark acoustic oscillations, and SIDM with a self-consistently chosen velocity-dependent cross-section.
As the TNG ETHOS galaxies include more than the standard SIDM (which would use the CDM initial matter power spectrum), we will refer to them as ETHOS/SIDM.

\subsubsection{Do simulated UFDs match observations?}
While there are many different metrics for determining how well simulations are producing UFDs (e.g., velocity dispersions, metallicity distribution functions), here, we choose to examine agreement in the $M_V-r_h$ plane. 
From Figure \ref{fig:rhMV}, it can be seen that the \cite{Jeon2021a,Jeon2021b} simulations (higher resolution, magenta x's) produce the highest number of UFD analogs that overlap with observed MW UFDs. EDGE \citep{agertz2020,prgomet2022}, FIRE-2 \citep{jahn2022,wetzel2023} using both CDM and WDM \citep{bozek2019}, GEAR \citep{revaz2018,sanati2023}, \cite{maccio2017,maccio2019}, and the Marvel-ous Dwarfs \citep{munshi2021} groups have also produced some UFDs that match the bright end of the observed population. 

The fact that observed UFD properties are able to be matched by high resolution simulations of isolated UFDs could suggest that environmental effects from the MW have not played a critical role in their size evolution. Additionally, the LMC-associated UFDs are matched by these isolated simulations. 
Their more recent infall ($\sim$2 Gyr ago), paired with their lack of differences in the $M_V-r_h$ space from the non-LMC-associated UFDs, further supports that the presence of the MW has had little overall consequence on the bulk of the current UFD size distribution.

Interestingly, in a study of the extended stellar populations of UFDs (i.e., member stars beyond four times $r_h$), \cite{tau2024} found that at least 10 satellites with external populations had a wide range of magnitude and pericenter distances. This too suggests that interactions with the MW are $not$ what dominates the distribution of UFD stars that we observe today.

Despite the overlap of simulations with the brighter UFDs, most simulations are unable to produce analogs of the most compact observed UFDs \citep[$r_h\sim30$ pc; e.g.,][]{revaz2023}. Rather, they are creating UFDs with half-light radii up to an order of magnitude larger than observed. 
\cite{revaz2023} speculates that this could be due to resolution limits, spurious numerical heating, or improper feedback prescriptions, although the widespread nature of the pattern suggests there is no singular factor. In their own work using DM-only simulations, \cite{revaz2023} found that mini-halo mergers played a large role in the expansion of the simulated UFD sizes.

Conversely, it could be said that simulations are producing galaxies that are more diffuse than currently known observationally. This disparity can be revisited as new observatories come online and we expand our knowledge of the low surface-brightness universe. Discoveries of larger, low surface-brightness UFDs could resolve the current discrepancy.

The simulations whose data we show in Figure \ref{fig:rhMV} have a wide range of mass resolutions and distinct implementations of subgrid physics, which are necessary to form galaxies in a cosmological simulation, as the range of scales (e.g., individual supernovae to Mpc-sized volumes) would be too computationally expensive otherwise. Despite using different combinations of free parameters, many simulation groups are still able to produce UFD analogs that fall within the same $M_V-r_h$ space. 
For example, simulated UFDs with $r_h$ between $\sim100-300$ pc and $M_V\leq-4$ are produced by the DC Justice League \citep{Applebaum2021}, EDGE \citep{agertz2020,orkney2021,rey2022}, FIRE-2 \citep{jahn2022,wetzel2023}, GEAR \citep{sanati2023}, \cite{Jeon2017}, \cite{maccio2017,maccio2019}, and the Marvel-ous Dwarfs \citep{munshi2021}. 

The consistently most compact UFDs ($r_h<100$ pc) come from \cite{Jeon2021a,Jeon2021b}. Paired with their low luminosities ($M_V \geq -4$), these simulations are quite successful in reproducing the observed distribution of the and LMC- and non-LMC-associated UFDs in the magnitude-size plane. These high-resolution simulations were run with a customized version of \textsc{gadget} \citep{springel2005} which used smoothed particle hydrodynamics with a gas particle mass of 63 $M_\odot$. A variable softening length was used for the gas particles with a minimum length of $\sim3$ pc. Individual Population III stars could be born out of pristine gas, while Population II stars were formed out of metal-enriched gas in complexes no less massive than 500 $M_\odot$ to emulate the birth of stars out of molecular clouds.

The UFDs from the EDGE simulations \citep{agertz2020,rey2020,orkney2021,prgomet2022} are slightly larger and more luminous than those of \cite{Jeon2021a,Jeon2021b}, but still fainter and more compact than many of the other simulated galaxies. 
Their properties relative to the \cite{Jeon2021a,Jeon2021b} galaxies are reasonable as the EDGE UFDs inhabited DM halos that were up to an order of magnitude more massive than those hosting the UFDs in \cite{Jeon2021a,Jeon2021b} (see Table \ref{tab:simtable}). The UFDs from the EDGE simulations track the observed magnitude-size relation well, overlapping with the LMC- and non-LMC-associated UFDs at the bright end ($M_V \leq -4$). These high-resolution simulations were run with the \textsc{ramses-rt} 
adaptive mesh refinement code \citep{Teyssier2002,Rosdahl2013,Rosdahl2015} at a variety of resolutions and with many variations in the baryonic physics.

In the UFD regime, besides the baryonic mass resolution, one could also consider the targeted halo mass at $z=0$ for simulations as a possible factor in the final galaxy size. For example, in a study specifically focused on reproducing the compactness of observed UFDs, \cite{revaz2023} used dark-matter-only (DMO) simulations to study how the initial size and later merger history of UFD building blocks affected the $z=0$ size of UFDs. They concluded that simulated UFDs with $r_h<30$ pc can form from single mini-halos with masses smaller than $4\times10^8 M_{\odot}$ at $z=6$ and initial sizes $<15$ pc. 

\cite{Jeon2021a,Jeon2021b} were able to hydrodynamically simulate UFDs on this size scale with halo masses $<2.5\times10^8 M_{\odot}$. A few other simulations had halo masses below this \citep[e.g.,][J.~Rose et al., 2024, in prep.]{Applebaum2021,rose2023}~but were unable to achieve the compactness. For \cite{Applebaum2021}, this could be due to the difficulty in simulating UFDs in addition to a MW-mass scale galaxy, while for TNG, their subgrid physics choices (as discussed further below) limited their minimum compactness.

The simulated galaxies that lie furthest from the observed sample and away from the bulk of the other simulations are the highest resolution FIRE-2 galaxies from \cite{wheeler2019}, which have very low surface brightnesses ($\mu_V\gtrsim30$ mag arcsec$^{-2}$), and the TNG galaxies in both the CDM and ETHOS/SIDM cosmologies \citep[][J.~Rose et al.~2024, in prep.]{rose2023}, which have very large sizes ($r_h\gtrsim10^3$ pc) (though it should be remembered that we are not expecting TNG to be producing true UFD analogs).

The FIRE-2 galaxies in this group have $\sim30 M_{\odot}$ baryonic particle resolution and were presented in \cite{wheeler2019}. 
Their larger sizes are not due to mass resolution or the force-softening scale, and the authors suggest that only the bright cores of UFDs are being captured by current observatories. When \cite{wheeler2019} applied a surface brightness detection limit of $<$30 mag arcsec$^{-2}$ and remeasured the apparent $r_h$ and $M_{\star}$, their simulated analogs moved into the same space as observed UFDs. 

As we were able to trace the morphologies of the dwarfs in our sample to an average surface brightness of $\mu_V=31.2$ mag arcsec$^{-2}$, we expect that our measurements were not significantly biased by our imaging depth (see also the discussion at the end of Section \ref{sec:fitting}). However, new observational facilities are set to come online in the next decade that should reach or exceed our photometric depth while covering the entirety of the dwarfs (e.g., the Vera C. Rubin Observatory/LSST, \citealt{ivezic2019}; and the Roman Space Telescope, \citealt{spergel2015}). As our HST fields are limited in their FOV, it will be interesting to see what these observatories may reveal about the extremities of these dwarf galaxies.

Of note, the \cite{wheeler2019} FIRE-2 galaxies employed the same subgrid physics model as other FIRE-2 galaxies (e.g., the \citealt{jahn2022}) in Figure \ref{fig:rhMV} that are less diffuse for their magnitude. The two differences between these sets of FIRE-2 simulations were their mass resolutions and the environments in which the galaxies were simulated. The \citeauthor{wheeler2019} study considered UFDs in isolated dwarf galaxy halos, while the work of \citeauthor{jahn2022} examined UFDs around an LMC-mass halo. Environment has previously been shown to affect DM halo properties such as concentration, where halos that formed earlier in cosmic time, when the Universe was more dense, have more highly concentrated density profiles \citep[e.g.,][]{wechsler2002}. Additionally, for galaxies with the same peak halo masses, those that formed in more dense environments tended to have higher stellar masses than those that were isolated \citep{christensen2023}. 

It has not yet been explored whether there is a correlation for simulated UFD size ($r_h$) or light profile concentration with relation to environment or formation time. As the target UFDs that we fit structural parameters for all had surface density profiles that were well described by exponential and/or Plummer profiles, they showed no signs of variation. A larger sample of observed UFDs with fewer FOV constraints could be used in tandem with simulations to investigate if the formation environment could have an impact on present-day morphological properties.

Focusing now on the TNG simulations, both sets of galaxies occupy similar spaces in the $M_V-r_h$ plane, suggesting that the different DM models might not have a large effect on observable size. We also see that they tend to have larger $r_h$ values, especially for the magnitude space that they occupy. This is due to how they implement their SF and pressurization of the multiphase ISM \citep[e.g.,][]{pillepich2018}. 

The IllustrisTNG model is unable to resolve SNe and uses a pressurized equation of state that does not allow the gas to cool below 10$^4$~K, thus prohibiting collapse on smaller physical scales.
As such, there is a limit to how compact the TNG dwarf galaxies can be, motivating the need for simulations based on more resolved subgrid physics 
with explicit treatments of the ISM to explore alternative DM models.

While the TNG simulated galaxies shown in Figure \ref{fig:rhMV} should not be used for observational predictions, they are a useful marker for seeing what well-tested cosmological models produce at these scales. The similarity in sizes for the TNG galaxies produced using CDM versus ETHOS/SIDM could also suggest that the galaxy formation model is dominant over the DM prescription.

\subsubsection{Alternative dark matter models}

In Figure \ref{fig:rhMV}, there is a subset of points that come from simulations based on alternate DM models, such as WDM \citep[e.g.,][]{maccio2019,bozek2019} and SIDM \citep[e.g.,][J.~Rose et al.~2024, in prep.]{robles2017}. Within the $M_V-r_h$ plane, there are not extreme differences apparent in the faint dwarf analogs created with different DM models. Below, we present a brief overview of the more in-depth literature studies that have compared these DM models on mass scales relevant to UFDs.

To leverage DMO simulations, we can first characterize any strong differences present between the DM models without the presence of baryons. Then, we can see if those differences persist in the fully hydrodynamic runs, or if the addition of baryons causes observational degeneracies. Comparisons of cold DMO simulations to observations of dwarf spiral galaxies previously led to the ``core-cusp'' problem, where measured rotation curves implied the presence of cores, rather than the predicted steep density profiles ($\rho\propto r^{-1}$) \citep[e.g.,][]{flores1994,moore1994}. While there have since been explanations for this difference consistent with CDM \citep[such as time-varying gravitational potentials, see e.g.,][]{pontzen2012,dutton2016,orkney2021}, alternative DM models have also been proposed to address the ``core-cusp'' and other challenges that $\Lambda$CDM seemingly faced on small ($<1$ Mpc) scales \citep[see][for a review]{bullock2017}. 

For example, SIDM-only simulations have been able to produce dwarf halos with constant-density cores ($r_c\sim1$ kpc) \citep[e.g.,][]{burkert2000,dave2001,rocha2013,robles2017}. WDM-only simulations have also produced less centrally dense halos \citep[e.g.,][]{lovell2012,shao2013}, although some simulations predict cores closer to $\sim10$ pc \citep{maccio2012}, which would be insufficient for resolving the core-cusp problem. These three DM models thus have distinct signatures present when there are no baryons involved. As we rely on observations of baryonic matter for science, however, we must see what effect the addition of baryons has on these simulations.

Here, we focus on baryonic simulations that used fixed subgrid physics for both CDM and alternative DM runs.\footnote{When discussing the simulations, unless specified as DM-only, we are referring to the baryonic simulations based on the DM models.}
In comparing UFD analogs in CDM versus SIDM cosmologies, both FIRE-2 \citep{robles2017} and TNG (ETHOS/SIDM) (J.~Rose et al.~2024, in prep.) found similar $M_{\star}-r_h$ relations between the two models. Additionally, cores were difficult to form in CDM, with only the most massive halo forming a core in the FIRE-2 runs and no halos forming cores in the TNG (CDM) runs \citep{rose2023}. The SIDM halos did form cores, although the UFD core sizes and profiles were not significantly different from those formed in the SIDM-only runs. This suggests that the baryons had minimal effects on the DM structure of the dwarf galaxies considered. \cite{robles2017} suggested that the observation of field dwarf with a DM core size similar to its $r_h$ and a stellar mass $\lesssim3\times10^6 M_{\odot}$ could support the need for a new DM paradigm.

For the simulated dwarfs in CDM versus WDM, \cite{bozek2019} (using FIRE-2 subgrid physics) and \cite{maccio2019} \citep[using Numerical Investigation of a Hundred Astrophysical Objects (NIHAO) subgrid models;][]{wang2015} both found that the WDM galaxies formed the bulk of their stellar mass in the past $\sim$4 Gyr. \cite{bozek2019} proposed that observations of UFDs with more than 80\% of their stellar mass formed in the last 4 Gyr would support the WDM model. 

Both groups also found that in general, the final WDM density profiles of halos hosting galaxies were less centrally dense than the halos in CDM, although baryonic feedback processes had a greater effect overall in lowering the central densities in the CDM runs; greater stellar mass and star formation efficiency were also correlated with greater central density reduction. \cite{maccio2019} additionally predicted WDM halos hosting galaxies with $M_{\star}<10^6$ would retain their cuspy profiles. Thus, the authors suggested that the unambiguous detection of a core in a UFD would require a more drastic change to the standard model. \cite{bozek2019} used the WDM resonantly-produced sterile neutrino (RPSN) particle model while \cite{maccio2019} chose a WDM particle mass of 3 keV, however both concluded that observables such as galaxy mass, shape, size, and velocity dispersion would not be useful in distinguishing between a CDM or WDM universe.
These two studies produced very few simulated galaxies with $M_V>=-10$, however, so it would be informative to have a larger sample size of galaxies from WDM simulations specifically targeting the UFD-mass regime.

Finally, \cite{fitts2019}\footnote{The simulated data unique to this study are not in Figure \ref{fig:rhMV} as they were not available.} compared the effects of FIRE-2 baryonic physics on WDM, SIDM, and CDM, as well as two models combining SIDM and WDM. The authors found that the simulated dwarfs across their models had similar stellar half-mass radii to observed dwarf galaxies in the Local Field, and that the $r_{1/2}-M_{\star}$ relation (where $r_{1/2}$ is the 3D half-mass radius) did not show different trends based on the DM model used. Similarly to the comparisons above, \cite{fitts2019} found that the addition of baryonic physics made the DM density profile differences that were evident in the DMO-simulated halos less apparent. 

In their chosen comparison plane of $V_{1/2}$ (the circular velocity at $r_{1/2}$) versus $r_{1/2}$, all five models (CDM, SIDM, WDM, and two SIDM+WDM) yielded similar results that agreed with observations. However, the alternate DM runs produced dwarfs with lower central densities than those in CDM for $M_{\star}\sim10^6 M_{\odot}$. \cite{fitts2019} suggested that measurements of circular velocities for the inner regions ($<400$ pc) of less massive dwarfs could be a strong discriminator for CDM versus an alternative DM. The authors concluded that a larger sample size of simulated dwarfs in this mass regime will be needed for a more thorough exploration. 

The studies discussed above used the FIRE-2, IllustrisTNG, and NIHAO subgrid models, but it would be informative to see how the baryonic physics implementations of other simulation groups interact with alternative DM models on faint dwarf galaxy mass scales. These could provide additional observational probes for differentiating between the proposed natures of DM.

\subsubsection{The low surface-brightness future}
As the next generation of observing facilities come online (e.g., Euclid, \citealt{racca2016}; Vera C. Rubin Observatory/LSST, \citealt{ivezic2019}; Roman Space Telescope, \citealt{spergel2015}), we can expect even more low surface-brightness stellar populations to be discovered. For example, \cite{mutluPakdil2021} simulated the expected performance of the Subaru Hyper Suprime-Cam \citep{miyazaki2018} and Rubin, forecasting that we should obtain a ``secure census'' of galaxies out to 1.5 Mpc with $M_V\lesssim-7$ with $\mu_{V,0}\sim30$ mag arcsec$^{-2}$. Similar investigations for resolved stellar streams with Roman \citep{pearson2019} and ultra-diffuse galaxies (UDGs) with LSST depth \citep{newton2023} have also been conducted, and the authors are optimistic about the populations awaiting discovery within our Local Volume. \cite{pearson2019} predicted that we will be able to detect resolved stellar streams from globular clusters out to $\sim$3.5 Mpc, and \cite{newton2023} predicted that there are $\sim$12 isolated UDGs within 2.5 Mpc of the Local Group that should be observable in future deep surveys.

As an additional note, from examining this section of the $M_V-r_h$ plane, there appears to be a deficit in the number of observed MW satellites with an $M_V$ between $-6$ and $-8$. 
This is the same feature in the luminosity function that \cite{bose2018} suggested may represent the dividing line between dwarfs that were quenched by reionization and those that were able to continue forming stars at later times.  
\cite{manwadkar2022} also highlight this ``break'' in their GRUMPY semi-analytic model and speculated that as the census of Local Volume satellite galaxies grows more complete, we can probe reionization by further characterizing the abundance of UFDs and how sensitive they must have been to the ionizing radiation from that epoch.

We will likely need a greater sample size to determine if this deficit is anything besides a random fluctuation.
Still, we might expect at least one MW satellite to be discovered with $100 < r_h < 400$ pc, as it would have a surface brightness above current and projected detection limits. Many hydrodynamic simulations have been able to produce galaxies in this magnitude range, and while there have been other Local Group galaxies discovered in that $M_V-r_h$ parameter space \citep[e.g.,][]{mcConnachie2012}, most of the currently known MW UFDs seem to be fainter than $M_V=-6$. 

There are further details to be studied concerning the faint and low surface-brightness populations that we have recently discovered. For example, there are still several stellar systems (e.g., Hor~II, Hya~II, Phe~II, Tri~II) whose status as galaxies (rather than star clusters) has yet to be confirmed via spectroscopy. The \cite{Baumgardt2022} mass segregation technique has provided an alternative, however, and concluded that Hor~II, Phe~II, and Tri~II are likely dwarf galaxies. 
The \cite{fu2023} HST photometric metallicity study has also offered support via metallicity dispersion values for the classification of Hya~II and Phe~II as galaxies.
The targets that we have labeled ``ultra-faint clusters'' also merit further studies, as not all have been conclusively ruled out of being DM-dominated \citep[e.g.,][]{cerny2021a,cerny2021b,cerny2023}. Additionally, studies of faint dwarf stellar halos \citep{jensen2023} and UFD stars at distant radii \citep[e.g., Tuc~II;][]{chiti2021} suggest that deeper imaging of the fields \textit{around} UFDs could reveal more about previous tidal interactions or whether they formed through the merger of smaller stellar systems \citep[e.g.,][]{goater2023}. The depth and areal coverage of the surveys to come will be critical in our characterization and understanding of the low surface-brightness universe.

\section{SUMMARY AND CONCLUSIONS} \label{sec:conc}

In this paper, we have presented new PSF photometry for ten targets from the HST Treasury GO-14734 (PI: N. Kallivayalil): Grus~II, Horologium~I, Horologium~II, Hydra~II, Phoenix~II, Reticulum~II, Sagittarius~II, Triangulum~II, Tucana~II, and Tucana~IV.
We have tested the ability to obtain structural parameter and $M_V$ measurements from HST imaging with limited fields-of-view and recovered $a_h$ measurements within $1\sigma$ of recent literature values for $\sim$40\% of the cases. 
While complications in the image data made these measurements difficult for several objects, all of our $a_h$ measurements agree within $3\sigma$ of published values.

We have shown that there is no systematic difference in the $M_V-r_h$ relationship for UFDs associated with the LMC versus not. 
We examined the current state of dwarf galaxy simulations and found that several groups have been able to create analogs that are broadly consistent with the observed UFDs in the magnitude-size plane.
We also considered whether \textit{where} observed UFDs fall in the $M_V-r_h$ plane would be useful in constraining the nature of DM. 
In comparing simulated galaxies from SIDM \citep[][J.~Rose et al.~2024, in prep.]{robles2017}, WDM \citep{bozek2019,maccio2019}, and CDM (see Table \ref{tab:simtable} for full citations) in the $M_V-r_h$ space, we found no consistent trends based on the DM cosmology. 

As we found no clear discriminator that would support one DM cosmology over another in the $M_V-r_h$ plane, we reviewed and compared the predicted observables from the literature.
From examining the results and conclusions of alternative DM studies from different simulation groups \citep[though see][for an outstanding comparison within the FIRE-2 group]{fitts2019}, two clear observables that would strongly support a non-CDM cosmology are 1) a UFD SFH with $\gtrsim80$\% of its stellar mass formed in the last 4 Gyr, which would agree with predictions from multiple WDM simulations \citep{bozek2019,maccio2019}, and 2) the presence of a core in a UFD with $M_{\star}\lesssim10^6 M_{\odot}$, as current simulations suggest that UFDs of that mass would retain their cusps \citep{robles2017,maccio2019,fitts2019}. 
Nonetheless, current hydrodyanmical simulations have generally been presenting baryonic solutions for the classical issues that CDM has faced on small scales \citep[e.g.,][]{sales2022} and a paradigm shift in the near-future would require substantial evidence.

Statistically robust samples of simulated dwarfs produced across different DM models will be necessary to more thoroughly explore the impact that the underlying DM cosmology has on tangible properties. Additionally, as more simulation groups explore alternative DM models, we can see if those, in combination with their subgrid physics implementations, suggest different observable relations that could be used to constrain DM properties. Finally, as new observing facilities come online and we begin to go beyond current survey depth limits, we will have more opportunities to discover and further characterize stellar populations in this low surface-brightness regime.

\begin{sidewaystable*}
\begin{center}
\begin{tabular}{llllll}
\multicolumn{6}{c}{\textbf{Horologium~I}}\\ 
\multicolumn{1}{l}{Parameter} & \cite{bechtol2015}          & \cite{koposov2015b}$^a$      & \cite{munoz2018}   & \cite{moskowitz2020}$^b$    & \multicolumn{1}{l}{This Work}               \\ \toprule
$M_V$ & $-3.5 \pm 0.3$ & $-3.4 \pm 0.1$ & $-3.55$ $\pm0.56$ & \multicolumn{1}{l}{...} & $-3.4$ $\pm0.2$ \\\midrule
\multicolumn{6}{c}{\textit{Exponential}}\\
RA (deg) & \multicolumn{1}{l}{...}  & 43.8820 & 43.8813 $\pm25.65\arcsec$ &... & 43.8759 $\pm3\arcsec$   \\
DEC (deg) & \multicolumn{1}{l}{...} & $-54.1188$ & $-54.1160$  $\pm20.2\arcsec$ &... & $-54.1174$ $\pm3\arcsec$  \\
$\theta_{\mathrm{exp}}$ (deg)  & \multicolumn{1}{l}{...} & Unconstrained      & 53 $\pm 27$ &... & 44 $\pm7$  \\
$\epsilon_{\mathrm{exp}}$  & \multicolumn{1}{l}{...} & $<$0.28 & 0.32 $\pm0.13$ &... & 0.26 $\pm0.05$ \\
$a_{h,\mathrm{exp}}$ (arcmin)          & \multicolumn{1}{l}{...}    & 1.41$^{+0.29}_{-0.13}$ & 1.71 $\pm0.37$ &... & 1.70$^{+0.16}_{-0.15}$ \\
$a_{h,\mathrm{exp}}$ (pc)          & \multicolumn{1}{l}{...}    & 33$^{+6.8}_{-4.8}$        & 39.3 $\pm8.5$ &... & 39.2$^{+5.2}_{-4.7}$ \\
\midrule
\multicolumn{6}{c}{\textit{Plummer}}\\
RA (deg) & 43.87 & \multicolumn{1}{l}{...} & 43.8813 $\pm25.65\arcsec$  & 43.882 & 43.8755 $\pm3\arcsec$\\
DEC (deg) & $-54.11$ & \multicolumn{1}{l}{...} & $-54.1160$ $\pm20.2\arcsec$ & $-54.119$ & $-54.1174$ $\pm3\arcsec$\\
$\theta_{p}$ (deg)            & Unconstrained   & \multicolumn{1}{l}{...}  & 57 $\pm25$ &   66.59$^{+13.41}_{-13.65}$  &  44 $\pm6$ \\
$\epsilon_{p}$   & Unconstrained  & \multicolumn{1}{l}{...}  &  0.27 $\pm0.13$ & 0.14$^{+0.06}_{-0.05}$ & 0.27 $\pm0.05$ \\
$a_{h,\mathrm{p}}$ (arcmin)            & 2.4$^{+3.0}_{-1.2}$   & \multicolumn{1}{l}{...}  &  1.59 $\pm0.31$ &    1.67$\pm0.10$ & 1.61 $\pm0.13$    \\ 
$a_{h,\mathrm{p}}$(pc)      & 60$^{+76}_{-30}$   & \multicolumn{1}{l}{...}   & 36.5 $\pm7.1$      & 38.6$^{+4.6}_{-3.9}$ &  37.3$^{+4.2}_{-4.7}$\\
\bottomrule
\end{tabular}
\end{center}
\caption{Compilation of past literature measurements for Horologium~I, which include fits from both the exponential and Plummer models. Our parameters are quoted from the maximum-likelihood estimation, with the listed uncertainties corresponding to the 84th and 16th percentiles around the median of the distribution.}
\label{tab:hor1}
\begin{FlushLeft}
\footnotesize{$^a$The values reported as $a_h$ in arcminutes here are the values given as $r_{\mathrm{maj}}$ in \cite{koposov2015b} Table 1 multiplied by 1.68, as their $r_{\mathrm{maj}}$ corresponded to the exponential scale length $r_e$.}\\
\footnotesize{$^b$The values reported as $a_h$ in arcminutes here are the values given as $R_{h,1c}$ divided by $\sqrt{1-\epsilon_{1c}}$ in their Table 3, as their $R_h$ is what they refer to as the ``circularized projected half-light radii'' equal to $a_h\sqrt{1-\epsilon_{1c}}$. \cite{moskowitz2020} do not report fitted RA and DEC values, thus we list the values given in their Table 2.}
\end{FlushLeft}
\end{sidewaystable*}

\begin{table*}
\begin{center}
\begin{tabular}{llll}
\multicolumn{4}{c}{\textbf{Horologium~II}}\\ 
\multicolumn{1}{l}{Parameter} & \cite{kim2015}  & \cite{moskowitz2020}$^c$  &  \multicolumn{1}{l}{This Work}      \\ \toprule
$M_V$ & $-2.6^{+0.2}_{-0.3}$ &  \multicolumn{1}{l}{...} & $-2.1 \pm0.2$
\\\midrule
\multicolumn{4}{c}{\textit{Exponential}}\\
RA (deg) & 49.1337  $\pm5\arcsec$  &  \multicolumn{1}{l}{...}  & 49.1315 $\pm4\arcsec$  \\
DEC (deg) & $-50.0180$ $\pm5\arcsec$ &  \multicolumn{1}{l}{...} &  $-50.0088$ $\pm3\arcsec$\\
$\theta_{\mathrm{exp}}$ (deg) & 127 $\pm11$ &  \multicolumn{1}{l}{...} &  103$^{+12}_{-14}$\\
$\epsilon_{\mathrm{exp}}$ & 0.52$^{+0.13}_{-0.17}$ & \multicolumn{1}{l}{...} &  0.23$^{+0.07}_{-0.08}$\\
$a_{h,\mathrm{exp}}$ (arcmin)  & 2.09$^{+0.44}_{-0.41}$ & \multicolumn{1}{l}{...} & 1.63 $\pm0.18$\\
$a_{h,\mathrm{exp}}$ (pc) & 47 $\pm10$ & \multicolumn{1}{l}{...} & 36.9$^{+5.1}_{-5.4}$\\
\midrule
\multicolumn{4}{c}{\textit{Plummer}}\\
RA (deg) & \multicolumn{1}{l}{...} & 49.134 & 49.1310 $\pm4\arcsec$\\
DEC (deg) & \multicolumn{1}{l}{...} & $-50.0181$ & $-50.0090$ $\pm3\arcsec$ \\
$\theta_{\mathrm{p}}$ (deg) & \multicolumn{1}{l}{...} & 279.71$^{+8.25}_{-189.75}$ & 103$^{+11}_{-14}$\\
$\epsilon_{\mathrm{p}}$ & \multicolumn{1}{l}{...} & 0.40 $\pm0.14$ & 0.23$^{+0.07}_{-0.08}$\\
$a_{h,\mathrm{p}}$ (arcmin)  & \multicolumn{1}{l}{...} & 2.54$^{+0.43}_{-0.53}$ & 1.69$^{+0.18}_{-0.17}$\\
$a_{h,\mathrm{p}}$ (pc) & \multicolumn{1}{l}{...} & 58.1$^{+13.8}_{-10.7}$ & 38.4$^{+4.9}_{-5.5}$\\
\bottomrule
\end{tabular}
\caption{The same as Table \ref{tab:hor1}, for Horologium~II.}
\label{tab:hor2}
\begin{FlushLeft}
\footnotesize{$^c$The values reported as $a_h$ in arcminutes here are the values given as $R_{h,1c}$ divided by $\sqrt{1-\epsilon_{1c}}$ in their Table 3, as their $R_h$ is what they refer to as the ``circularized projected half-light radii'' equal to $a_h\sqrt{1-\epsilon_{1c}}$. \cite{moskowitz2020} do not report fitted RA and DEC values, thus we list the values given in their Table 2.}
\end{FlushLeft}
\end{center}
\end{table*}

\begin{table*}
\centering
\begin{tabular}{llll}
\multicolumn{4}{c}{\textbf{Hydra~II}}\\ 
\multicolumn{1}{l}{Parameter} & \cite{martin2015}  & \cite{munoz2018}  & \multicolumn{1}{l}{This Work}  \\ \toprule
$M_V$ & $-4.8$ $\pm0.3$ & $-4.60$ $\pm0.37$ & $-4.6^{+0.2}_{-0.3}$ \\\midrule
\multicolumn{4}{c}{\textit{Exponential}}\\
RA (deg) & 185.4254 & 185.4251 $\pm13.65\arcsec$ & 185.4279$\pm4\arcsec$\\ 
DEC (deg) & $-31.9853$ & $-31.9860$ $\pm13.7\arcsec$ & $-31.9728^{+7}_{-6}\arcsec$\\ 
$\theta_{\mathrm{exp}}$ (deg) & 28$^{+40}_{-35}$ & 13 $\pm28$ & $-10^{+9}_{-10}$\\
$\epsilon_{\mathrm{exp}}$ & 0.01$^{+0.19}_{-0.01}$ & 0.25 $\pm0.16$ & 0.30$^{+0.08}_{-0.09}$\\ 
$a_{h,\mathrm{exp}}$ (arcmin)  & 1.7$^{+0.3}_{-0.2}$ & 1.65 $\pm0.39$ & 2.70$^{+0.41}_{-0.38}$\\ 
$a_{h,\mathrm{exp}}$ (pc) & 68 $\pm 11$ & 64.3 $\pm15.2$ & 105.6$^{+17.9}_{-16.1}$\\ 
\midrule
\multicolumn{4}{c}{\textit{Plummer}}\\
RA (deg) & \multicolumn{1}{l}{...} & 185.4251 $\pm13.65\arcsec$ & 185.4286$^{+4}_{-3}\arcsec$\\
DEC (deg) & \multicolumn{1}{l}{...} & $-31.9860$ $\pm13.7\arcsec$ & $-31.9728^{+9}_{-7}\arcsec$\\
$\theta_{p}$ (deg) & \multicolumn{1}{l}{...} & 16 $\pm25$ & $-9^{+8}_{-9}$\\
$\epsilon_{p}$ & \multicolumn{1}{l}{...} & 0.24 $\pm0.16$ & 0.30$\pm0.09$\\
$a_{h,\mathrm{p}}$(arcmin)  & \multicolumn{1}{l}{...} & 1.52 $\pm0.28$ & 2.13$^{+0.42}_{-0.35}$\\
$a_{h,\mathrm{p}}$(pc) & \multicolumn{1}{l}{...} & 59.2 $\pm10.9$ & 83.3$^{+17.1}_{-15.1}$\\
\bottomrule
\end{tabular}
\caption{The same as Table \ref{tab:hor1}, for Hydra~II.}
\label{tab:hya2}
\end{table*}

\begin{sidewaystable*}
\centering
\begin{tabular}{lllllll}
\multicolumn{7}{c}{\textbf{Phoenix~II}}\\ 
\multicolumn{1}{l}{Parameter} & \cite{bechtol2015}$^d$  & \cite{koposov2015b}$^e$  & \cite{munoz2018}   & \cite{mutluPakdil2018}  & \cite{moskowitz2020}$^f$  & \multicolumn{1}{l}{This Work} \\ \toprule
$M_V$ & $-3.7$ $\pm0.4$ & $-2.8$ $\pm0.2$ & $-3.30$ $\pm0.63$ & $-2.7$ $\pm0.4$ & \multicolumn{1}{l}{...} & -2.9$^{+0.2}_{-0.1}$ \\\midrule
\multicolumn{7}{c}{\textit{Exponential}}\\
RA (deg) & \multicolumn{1}{l}{...} & 354.9975 & 354.9960 $\pm13.5\arcsec$ & 354.9928 $\pm8.3\arcsec$ & \multicolumn{1}{l}{...} & 354.9922 $\pm3\arcsec$\\
DEC (deg) & \multicolumn{1}{l}{...} & $-54.4060$ & $-54.4115$ $\pm21.0\arcsec$ & $-54.4050$ $\pm5.7\arcsec$ & \multicolumn{1}{l}{...}  & $-54.4018$ $\pm3\arcsec$\\
$\theta_{\mathrm{exp}}$ (deg) & \multicolumn{1}{l}{...} & 164 $\pm54$ & $-19$ $\pm15$ & 156 $\pm13$ & \multicolumn{1}{l}{...} & $-34$ $\pm4$\\
$\epsilon_{\mathrm{exp}}$ & \multicolumn{1}{l}{...} & 0.47$^{+0.08}_{-0.29}$ & 0.62 $\pm0.19$  & 0.4 $\pm0.1$ & \multicolumn{1}{l}{...} & 0.44 $\pm0.6$\\
$a_{h,\mathrm{exp}}$ (arcmin)  & \multicolumn{1}{l}{...} & 1.38$^{+0.45}_{-0.20}$ & 1.60 $\pm0.33$ & 1.5 $\pm0.3$ & \multicolumn{1}{l}{...} & 1.58$^{+0.26}_{-0.23}$\\
$a_{h,\mathrm{exp}}$ (pc) & \multicolumn{1}{l}{...} & 34.3$^{+10.7}_{-6.4}$ & 38.6 $\pm8.0$  & 37 $\pm6$ & \multicolumn{1}{l}{...} & 43.0$^{+8}_{-7}$ \\
\midrule
\multicolumn{7}{c}{\textit{Plummer}}\\
RA (deg) & 354.99 & \multicolumn{1}{l}{...} & 354.9960 $\pm13.5\arcsec$ & \multicolumn{1}{l}{...} & 354.998 & 354.9919 $\pm3\arcsec$ \\
DEC (deg) & $-54.41$ & \multicolumn{1}{l}{...} & $-54.4115$ $\pm21.0\arcsec$ & \multicolumn{1}{l}{...} & $-54.406$ & $-54.4019$ $\pm3\arcsec$\\
$\theta_{p}$ (deg) & Unconstrained & \multicolumn{1}{l}{...} & $-20$ $\pm18$  & \multicolumn{1}{l}{...} & 285.28$^{+15.31}_{-14.54}$ & $-33$ $\pm5$\\
$\epsilon_{p}$ & Unconstrained & \multicolumn{1}{l}{...} & 0.67 $\pm0.22$ & \multicolumn{1}{l}{...} & 0.27$^{+0.11}_{-0.10}$ & 0.44 $\pm0.06$\\
$a_{h,\mathrm{p}}$(arcmin)  & 1.2$^{+1.2}_{-1.2}$ & \multicolumn{1}{l}{...} & 1.49 $\pm0.53$ & \multicolumn{1}{l}{...} & 1.85$^{+0.30}_{-0.33}$ & 1.50$^{+0.20}_{-0.17}$\\
$a_{h,\mathrm{p}}$(pc) & 33$^{+20}_{-11}$ & \multicolumn{1}{l}{...} & 36.0 $\pm12.8$ & \multicolumn{1}{l}{...} & 45.1$^{+9.5}_{-7.8}$ & 40.8$^{+5.5}_{-6.2}$\\
\bottomrule
\end{tabular}
\caption{The same as \ref{tab:hor1}, for Phoenix~II.}
\label{tab:phe2}
\begin{FlushLeft}
\footnotesize{$^d$The values reported as $a_h$ in arcminutes here are the values given as $r_h$ (deg) in \cite{bechtol2015} Table 1, multiplied by 60. The values reported as $a_h$ in parsecs here are the values given as $r_{1/2}$ (pc) in \cite{bechtol2015} Table 2.}\\
\footnotesize{$^e$The values reported as $a_h$ in arcminutes here are the values given as $r_{\mathrm{maj}}$ in \cite{koposov2015b} Table 1 multiplied by 1.68, as their $r_{\mathrm{maj}}$ corresponded to the exponential scale length $r_e$.}\\
\footnotesize{$^f$The values reported as $a_h$ in arcminutes here are the values given as $R_{h,1c}$ divided by $\sqrt{1-\epsilon_{1c}}$ in their Table 3, as their $R_h$ is what they refer to as the ``circularized projected half-light radii'' equal to $a_h\sqrt{1-\epsilon_{1c}}$. \cite{moskowitz2020} do not report fitted RA and DEC values, thus we list the values given in their Table 2.}
\end{FlushLeft}
\end{sidewaystable*}

\begin{sidewaystable*}
\begin{tabular}{llllll}
\multicolumn{6}{c}{\textbf{Sagittarius~II}}\\ 
\multicolumn{1}{l}{Parameter} & \cite{laevens2015b}  & \cite{mutluPakdil2018}   & \cite{longeard2020}   &\cite{moskowitz2020}$^g$  & \multicolumn{1}{l}{This Work}           \\ \toprule
$M_V$ & $-5.2$ $\pm0.4$ & $-4.2$ $\pm0.1$ & $-5.7$ $\pm0.1$ & \multicolumn{1}{l}{...} & $-5.3\pm0.2$ \\\midrule
\multicolumn{6}{c}{\textit{Exponential}}\\
RA (deg) & 298.1688 & 298.1647 $\pm3.0\arcsec$ & 298.16628 $\pm3.6\arcsec$ & \multicolumn{1}{l}{...} & 298.1664 $\pm2\arcsec$\\
DEC (deg) & $-22.0681$ & $-22.0651$ $\pm2.2\arcsec$ & $-22.89633$ $\pm3.6\arcsec$ & \multicolumn{1}{l}{...} & $-22.0641$ $\pm2\arcsec$\\
$\theta_{\mathrm{exp}}$ (deg) & 72$^{+28}_{-20}$ & Unconstrained & 103$^{+28}_{-17}$ & \multicolumn{1}{l}{...} & 98$^{+43}_{-30}$\\
$\epsilon_{\mathrm{exp}}$ & 0.23$^{+0.17}_{-0.23}$ & $<$0.1 & 0.0, $<$0.12 at 95\% CL & \multicolumn{1}{l}{...} & 0.03,$<0.09$ at 95\% CL\\
$a_{h,\mathrm{exp}}$ (arcmin)  & 2.0$^{+0.4}_{-0.3}$ & 1.6 $\pm0.1$ & 1.7 $\pm0.05$ & \multicolumn{1}{l}{...} & 1.94 $\pm0.08$\\
$a_{h,\mathrm{exp}}$ (pc) & 38$^{+8}_{-7}$ & 32 $\pm1$ & 35.5$^{+1.4}_{-1.2}$ & \multicolumn{1}{l}{...} & 39 $\pm4$\\
\midrule
\multicolumn{6}{c}{\textit{Plummer}}\\
RA (deg) & \multicolumn{1}{l}{...} & \multicolumn{1}{l}{...} & \multicolumn{1}{l}{...} & 298.169 & 298.1664 $\pm2\arcsec$\\
DEC (deg) & \multicolumn{1}{l}{...} & \multicolumn{1}{l}{...} & \multicolumn{1}{l}{...} & $-22.068$ & $-22.0642$ $\pm2\arcsec$\\
$\theta_{p}$ (deg) & \multicolumn{1}{l}{...} & \multicolumn{1}{l}{...} & \multicolumn{1}{l}{...} & 74.90$^{+14.69}_{-13.02}$ & 96$^{+50}_{-32}$\\
$\epsilon_{p}$ & \multicolumn{1}{l}{...} & \multicolumn{1}{l}{...} & \multicolumn{1}{l}{...} & 0.24$^{+0.10}_{-0.09}$ & 0.03,$<0.08$ at 95\% CL\\
$a_{h,\mathrm{p}}$(arcmin)  & \multicolumn{1}{l}{...} & \multicolumn{1}{l}{...} & \multicolumn{1}{l}{...} & 1.87$^{+0.19}_{-0.22}$ & 1.85$^{+0.07}_{-0.08}$\\
$a_{h,\mathrm{p}}$(pc) & \multicolumn{1}{l}{...} & \multicolumn{1}{l}{...} & \multicolumn{1}{l}{...} & 36.7$^{+5.1}_{-4.3}$ & 37 $\pm4$\\
\bottomrule
\end{tabular}
\caption{Same as Table \ref{tab:hor1}, for Sagittarius~II.}
\footnotesize{$^g$The values reported as $a_h$ in arcminutes here are the values given as $R_{h,1c}$ divided by $\sqrt{1-\epsilon_{1c}}$ in their Table 3, as their $R_h$ is what they refer to as the ``circularized projected half-light radii'' equal to $a_h\sqrt{1-\epsilon_{1c}}$. \cite{moskowitz2020} do not report fitted RA and DEC values, thus we list the values given in their Table 2.}
\label{tab:sag2}
\end{sidewaystable*}

\begin{sidewaystable*}
\centering
\begin{tabular}{llllll}
\multicolumn{6}{c}{\textbf{Triangulum~II}}\\ 
\multicolumn{1}{l}{Parameter} & \cite{laevens2015a}  & \cite{carlin2017}  &   \cite{munoz2018}  &\cite{moskowitz2020}$^h$ & \multicolumn{1}{l}{This Work}               \\ \toprule
$M_V$ & $-1.8$ $\pm0.5$ & $-1.2$ $\pm0.4$ & $-1.60$ $\pm0.76$ &\multicolumn{1}{l}{...} & $-1.4^{+0.3}_{-0.2}$\\\midrule
\multicolumn{6}{c}{\textit{Exponential}}\\
RA (deg) & 33.3225 & 33.3223 $\pm14.4\arcsec$ & 33.3252 $\pm14.55\arcsec$ &\multicolumn{1}{l}{...} & 33.3152 $\pm8\arcsec$\\
DEC (deg) & 36.1784 & 36.1719 $\pm9.7\arcsec$  & 36.1702 $\pm19.0\arcsec$ & \multicolumn{1}{l}{...} & 36.1704$^{+8}_{-9}\arcsec$\\
$\theta_{\mathrm{exp}}$ (deg) & 56$^{+16}_{-24}$ & 73 $\pm17$ & 28 $\pm19$  & \multicolumn{1}{l}{...} & 81$^{+30}_{-39}$ \\
$\epsilon_{\mathrm{exp}}$ & 0.21$^{+0.17}_{-0.21}$ & 0.3 $\pm0.1$ & 0.48 $\pm0.17$ & \multicolumn{1}{l}{...} & 0.15$^{+0.12}_{-0.10}$\\
$a_{h,\mathrm{exp}}$ (arcmin)  & 3.9$^{+1.1}_{-0.9}$ & 2.5 $\pm0.3$ & 2.34 $\pm0.58$ &\multicolumn{1}{l}{...} & 3.50$^{+0.59}_{-0.65}$\\
$a_{h,\mathrm{exp}}$ (pc) & 34$^{+9}_{-8}$ & 21 $\pm4$ & 20.4 $\pm5.1$ & \multicolumn{1}{l}{...} & 29.0$^{+5.2}_{-5.4}$\\
\midrule
\multicolumn{6}{c}{\textit{Plummer}}\\
RA (deg) & \multicolumn{1}{l}{...} & \multicolumn{1}{l}{...} & 33.3252 $\pm14.55\arcsec$ & 33.323 & 33.3155 $\pm9\arcsec$\\
DEC (deg) & \multicolumn{1}{l}{...} & \multicolumn{1}{l}{...} & 36.1702 $\pm19.0\arcsec$ & 36.1783 & 36.1691$^{+10}_{-12}\arcsec$\\
$\theta_{p}$ (deg) & \multicolumn{1}{l}{...} & \multicolumn{1}{l}{...} & 44 $\pm18$ & 300.72$^{+27.55}_{-222.65}$ & 79$^{+28}_{-38}$\\
$\epsilon_{p}$ & \multicolumn{1}{l}{...} & \multicolumn{1}{l}{...} & 0.46 $\pm0.16$ & 0.26$^{+0.13}_{-0.10}$ & 0.15$^{+0.12}_{-0.10}$\\
$a_{h,\mathrm{p}}$(arcmin)  & \multicolumn{1}{l}{...} & \multicolumn{1}{l}{...} & 1.99 $\pm0.49$ & 1.80$^{+0.30}_{-0.33}$ & 3.25$^{+0.51}_{-0.59}$\\
$a_{h,\mathrm{p}}$(pc) & \multicolumn{1}{l}{...} & \multicolumn{1}{l}{...} & 17.4 $\pm4.3$ & 16.0$^{+3.0}_{-2.7}$ & 27.0$^{+4.5}_{-4.9}$\\
\bottomrule
\end{tabular}
\caption{Same as Table \ref{tab:hor1}, for Triangulum~II, except our parameters are quoted from the median, with the listed uncertainties corresponding to the 84th and 16th percentiles around the median of the distribution.}
\begin{FlushLeft}
\footnotesize{$^h$The values reported as $a_h$ in arcminutes here are the values given as $R_{h,1c}$ divided by $\sqrt{1-\epsilon_{1c}}$ in their Table 3, as their $R_h$ is what they refer to as the ``circularized projected half-light radii'' equal to $a_h\sqrt{1-\epsilon_{1c}}$. \cite{moskowitz2020} do not report fitted RA and DEC values, thus we list the values given in their Table 2.}
\end{FlushLeft}
\label{tab:tri2}
\end{sidewaystable*}

\begin{sidewaystable*}
\begin{center}
\begin{tabular}{llll}
\multicolumn{4}{c}{\textbf{Grus~II}}\\ 
\multicolumn{1}{l}{Parameter} & \cite{drlica2015}  & \cite{moskowitz2020}$^i$  & \cite{simon2020}   \\ \toprule
$M_V$ &-3.9 $\pm0.22$ & \multicolumn{1}{l}{...} & -3.5 $\pm0.3$
\\\midrule
\multicolumn{4}{c}{\textit{Plummer}}\\
RA (deg) & 331.02 &  \multicolumn{1}{l}{...} & 331.025$^{32}_{29}$\\
DEC (deg) & $-46.44$ & $-46.44$ & $-46.422$ $\pm22$\\
$\theta_{\mathrm{p}}$ (deg) & Unconstrained & 67.05$^{+228.74}_{-26.25}$ & Unconstrained\\
$\epsilon_{\mathrm{p}}$ & $<0.2$ & 0.12$^{+0.06}_{-0.05}$ & $<0.21$\\
$a_{h,\mathrm{p}}$ (arcmin)  & 6.0$^{+0.9}_{-0.5}$ & 7.80$^{+0.56}_{-0.58}$ & 5.9 $\pm$0.5\\
$a_{h,\mathrm{p}}$ (pc) & 93 $\pm14$ & 121.5$^{15.23}_{-13.89}$ & 94 $\pm9$\\
\bottomrule
\end{tabular}
\end{center}
\caption{Compilation of past literature measurements for Grus~II from the Plummer model fits.}
\label{tab:gru2}
\footnotesize{$^i$The values reported as $a_h$ in arcminutes here are the values given as $R_{h,1c}$ divided by $\sqrt{1-\epsilon_{1c}}$ in their Table 3, as their $R_h$ is what they refer to as the ``circularized projected half-light radii'' equal to $a_h\sqrt{1-\epsilon_{1c}}$. \cite{moskowitz2020} do not report fitted RA and DEC values, however the literature values that they list in their Table 2 for this target do not match other studies. Their listed RA value does match up with other reported DEC values, so we do include that here.}
\end{sidewaystable*}

\begin{sidewaystable*}
\centering
\begin{tabular}{llllll}
\multicolumn{6}{c}{\textbf{Reticulum~II}}\\ 
\multicolumn{1}{l}{Parameter} & \cite{bechtol2015}$^j$  & \cite{koposov2015b}$^k$  & \cite{munoz2018}   & \cite{mutluPakdil2018}  & \cite{moskowitz2020}$^l$  \\ \toprule
$M_V$ & $-3.6$ $\pm0.1$ & $-2.7$ $0.1$ & $-3.88$ $\pm0.38$ & $-3.1$ $\pm0.1$ & \multicolumn{1}{l}{...} \\\midrule
\multicolumn{6}{c}{\textit{Exponential}}\\
RA (deg) & \multicolumn{1}{l}{...} & 53.9256 & 53.9203 $\pm24.45\arcsec$ & 53.9493 $\pm24.8\arcsec$ & \multicolumn{1}{l}{...} \\
DEC (deg) & \multicolumn{1}{l}{...} & $-54.0492$ & $-54.0513$ $\pm7.9\arcsec$ & $-54.0466$ $\pm9.1\arcsec$ & \multicolumn{1}{l}{...} \\
$\theta_{\mathrm{exp}}$ (deg) & \multicolumn{1}{l}{...} & 71 $\pm1$ & 62 $\pm2$ & 68 $\pm2$ & \multicolumn{1}{l}{...} \\
$\epsilon_{\mathrm{exp}}$ & \multicolumn{1}{l}{...} & 0.59$^{+0.02}_{-0.03}$ & 0.56 $\pm0.03$ & 0.6 $\pm0.1$ & \multicolumn{1}{l}{...} \\
$a_{h,\mathrm{exp}}$ (arcmin)  & \multicolumn{1}{l}{...} & 5.66$^{+0.39}_{-0.22}$ & 5.41 $\pm0.18$ & 6.3 $\pm0.4$ & \multicolumn{1}{l}{...} \\
$a_{h,\mathrm{exp}}$ (pc) & \multicolumn{1}{l}{...} & 50$^{+6}_{-5}$ & 47.2 $\pm1.6$ & 58 $\pm4$ & \multicolumn{1}{l}{...} \\
\midrule
\multicolumn{6}{c}{\textit{Plummer}}\\
RA (deg) & 53.92 & \multicolumn{1}{l}{...} & 53.9203 $\pm24.45\arcsec$ & \multicolumn{1}{l}{...} & 53.925\\
DEC (deg) & $-54.05$ & \multicolumn{1}{l}{...} & $-54.0513$ $\pm7.9\arcsec$ & \multicolumn{1}{l}{...} & $-54.049$\\
$\theta_{p}$ (deg) & 72 $\pm7$ & \multicolumn{1}{l}{...} & 70 $\pm2$ & \multicolumn{1}{l}{...} & 69.51 $\pm0.80$\\
$\epsilon_{p}$ & 0.6$^{+0.1}_{-0.2}$ & \multicolumn{1}{l}{...} & 0.58 $\pm0.02$ & \multicolumn{1}{l}{...} & 0.60 $\pm0.01$\\
$a_{h,\mathrm{p}}$(arcmin)  & 6.0 $\pm0.6$ & \multicolumn{1}{l}{...} & 5.52 $\pm0.19$ & \multicolumn{1}{l}{...} & 6.56 $\pm0.15$\\
$a_{h,\mathrm{p}}$(pc) & 55 $\pm5$ & \multicolumn{1}{l}{...} & 48.2 $\pm1.7$ & \multicolumn{1}{l}{...} & 57.9$^{+5.6}_{-5.3}$\\
\bottomrule
\end{tabular}
\caption{Same as Table \ref{tab:hor1}, for Reticulum~II.}
\label{tab:ret2}
\footnotesize{$^j$The values reported as $a_h$ in arcminutes here are the values given as $r_h$ (deg) in \cite{bechtol2015} Table 1, multiplied by 60. The values reported as $a_h$ in parsecs here are the values given as $r_{1/2}$ (pc) in \cite{bechtol2015} Table 2.}\\
\footnotesize{$^k$The values reported as $a_h$ in arcminutes here are the values given as $r_{\mathrm{maj}}$ in \cite{koposov2015b} Table 1 multiplied by 1.68, as their $r_{\mathrm{maj}}$ corresponded to the exponential scale length $r_e$.}\\
\footnotesize{$^l$The values reported as $a_h$ in arcminutes here are the values given as $R_{h,1c}$ divided by $\sqrt{1-\epsilon_{1c}}$ in their Table 3, as their $R_h$ is what they refer to as the ``circularized projected half-light radii'' equal to $a_h\sqrt{1-\epsilon_{1c}}$. \cite{moskowitz2020} do not report fitted RA and DEC values, thus we list the values given in their Table 2.}
\end{sidewaystable*}

\begin{table*}
\begin{center}
\begin{tabular}{llll}
\multicolumn{4}{c}{\textbf{Tucana~II}}\\ 
\multicolumn{1}{l}{Parameter} & \cite{bechtol2015}$^m$  & \cite{koposov2015b}$^n$ & \cite{moskowitz2020}$^o$ \\ \toprule
$M_V$ & $-3.9$ $\pm0.2$ & $-3.8$ $\pm0.1$ & \multicolumn{1}{l}{...}\\\midrule
\multicolumn{4}{c}{\textit{Exponential}}\\
RA (deg) & \multicolumn{1}{l}{...} & 342.9796 & \multicolumn{1}{l}{...}\\
DEC (deg) & \multicolumn{1}{l}{...} & $-58.5689$ & \multicolumn{1}{l}{...}\\
$\theta_{\mathrm{exp}}$ (deg) & \multicolumn{1}{l}{...} & 107 $\pm18$  & \multicolumn{1}{l}{...}\\
$\epsilon_{\mathrm{exp}}$ & \multicolumn{1}{l}{...} & 0.39$^{+0.10}_{-0.20}$ &  \multicolumn{1}{l}{...}\\
$a_{h,\mathrm{exp}}$ (arcmin)  & \multicolumn{1}{l}{...} & 12.89$^{+1.71}_{-1.98}$  & \multicolumn{1}{l}{...}\\
$a_{h,\mathrm{exp}}$ (pc) & \multicolumn{1}{l}{...} & 217$^{+40}_{-37}$  & \multicolumn{1}{l}{...}\\
\midrule
\multicolumn{4}{c}{\textit{Plummer}}\\
RA (deg) & 343.06 & \multicolumn{1}{l}{...} & 342.98\\
DEC (deg) & $-58.57$ & \multicolumn{1}{l}{...} & $-58.57$\\
$\theta_{p}$ (deg) & Unconstrained & \multicolumn{1}{l}{...} & 274.02$^{+6.81}_{-187.37}$\\
$\epsilon_{p}$ & Unconstrained & \multicolumn{1}{l}{...} & 0.34$^{+0.07}_{-0.08}$\\
$a_{h,\mathrm{p}}$(arcmin)  & 7.2 $\pm1.8$ & \multicolumn{1}{l}{...} & 13.48$^{+1.10}_{-1.20}$\\
$a_{h,\mathrm{p}}$(pc) & 120 $\pm30$ & \multicolumn{1}{l}{...} & 230.1$^{+31.0}_{-27.7}$\\
\hline
\end{tabular}
\end{center}
\caption{Same as Table \ref{tab:hor1}, for Tucana~II.}
\label{tab:tuc2}
\footnotesize{$^m$The values reported as $a_h$ in arcminutes here are the values given as $r_h$ (deg) in \cite{bechtol2015} Table 1, multiplied by 60. The values reported as $a_h$ in parsecs here are the values given as $r_{1/2}$ (pc) in \cite{bechtol2015} Table 2.}\\
\footnotesize{$^n$The values reported as $a_h$ in arcminutes here are the values given as $r_{\mathrm{maj}}$ in \cite{koposov2015b} Table 1 multiplied by 1.68, as their $r_{\mathrm{maj}}$ corresponded to the exponential scale length $r_e$.}\\
\footnotesize{$^o$The values reported as $a_h$ in arcminutes here are the values given as $R_{h,1c}$ divided by $\sqrt{1-\epsilon_{1c}}$ in their Table 3, as their $R_h$ is what they refer to as the ``circularized projected half-light radii'' equal to $a_h\sqrt{1-\epsilon_{1c}}$. \cite{moskowitz2020} do not report fitted RA and DEC values, thus we list the values given in their Table 2.}
\end{table*}

\begin{table*}
\begin{center}
\begin{tabular}{llll}
\multicolumn{4}{c}{\textbf{Tucana~IV}}\\ 
\multicolumn{1}{l}{Parameter} & \cite{drlica2015}  & \cite{moskowitz2020}$^p$  & \cite{simon2020}            \\ \toprule
$M_V$ & -3.5 $\pm0.28$ & \multicolumn{1}{l}{...} & -3.0$^{+0.3}_{-0.4}$
\\\midrule
\multicolumn{4}{c}{\textit{Plummer}}\\
RA (deg) & 0.73 & 0.73 & 0.717$^{+50}_{-76}$\\
DEC (deg) & $-60.85$ & $-60.85$ & $-60.830^{+36}_{-40}$\\
$\theta_{\mathrm{p}}$ (deg) & 11 $\pm9$ & 333.32$^{+9.26}_{-54.63}$ & 27$^{+9}_{-8}$\\
$\epsilon_{\mathrm{p}}$ & 0.4 $\pm0.1$ & 0.35$^{+0.07}_{-0.09}$ & 0.39$^{+0.07}_{-0.10}$\\
$a_{h,\mathrm{p}}$ (arcmin)  & 11.8$^{+2.2}_{-1.8}$ & 9.20$^{+0.96}_{-1.05}$ & 9.3$^{+1.4}_{-0.9}$\\
$a_{h,\mathrm{p}}$ (pc) & 167$^{+35}_{30}$ & 129.2$^{+19.3}_{-16.8}$ & 127$^{+22}_{-16}$\\
\bottomrule
\end{tabular}
\end{center}
\caption{Same as Table \ref{tab:gru2}, for Tucana~IV.}
\label{tab:tuc4}
\footnotesize{$^p$The values reported as $a_h$ in arcminutes here are the values given as $R_{h,1c}$ divided by $\sqrt{1-\epsilon_{1c}}$ in their Table 3, as their $R_h$ is what they refer to as the ``circularized projected half-light radii'' equal to $a_h\sqrt{1-\epsilon_{1c}}$. \cite{moskowitz2020} do not report fitted RA and DEC values, thus we list the values given in their Table 2.}
\end{table*}

\section*{Acknowledgments}
\begin{acknowledgments}
HR would like to thank Peter B. Stetson, Tom Brown, Taylor Hoyt, and Rachael Beaton for useful discussions about DAOPHOT. HR acknowledges support from the Jefferson Scholars Foundation Dissertation Year Fellowship, the Hilliard Family, and the Virginia Space Grant Consortium Graduate Research STEM Fellowship.
CTG acknowledges support from the Owens Family Foundation.
This work was based on observations with the NASA/ESA Hubble Space Telescope obtained from the Mikulski Archive for Space Telescopes at the Space Telescope Science Institute, which is operated by the Association of Universities for Research in Astronomy, Incorporated, under NASA
contract NAS5-26555. Support for HST GO-14734 was provided through a grant from the STScI under NASA contract NAS5-26555. This research has made use of NASA’s Astrophysics Data System.
\end{acknowledgments}
\noindent \textit{Facility}: {HST (ACS, WFC3)}

\noindent \textit{Software}: 
Aladin \citep{aladin};
Astrodrizzle \citep{Fructer2002,stsci2012,avila2015};
Astropy \citep{astropy:2013, astropy:2018};
corner.py \citep{corner};
DAOPHOT-II \citep{stetson1987,stetson1992};
dustmaps \citep{Green2018};
emcee \citep{ForemanMackey2013};
Jupyter Notebook \citep{soton403913};
Matplotlib \citep{Hunter:2007};
Numpy \citep{harris2020array};
photutils \citep{Bradley2020};
scikit-learn \citep{scikit-learn}
Scipy \citep{2020SciPy-NMeth};
stsynphot \citep{stsynphot};
synphot \citep{synphot}

\appendix
\section{Additional Corner Plots} \label{app:corner}

\begin{figure*}
     \centering
    \includegraphics[width=0.75\textwidth]{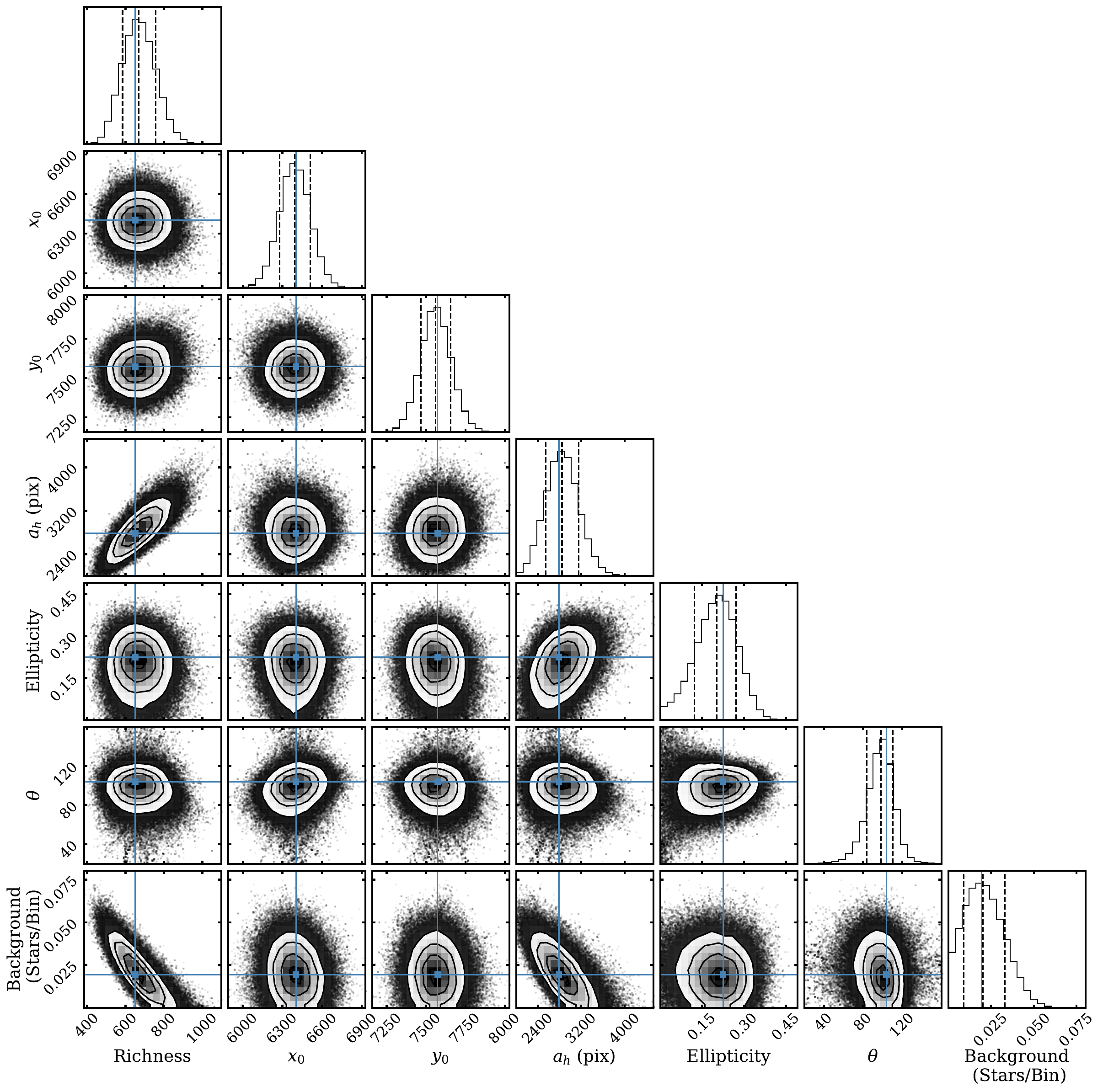}
    \caption{The posterior distributions of the 7-parameter structural fit for Hor~II using the exponential model. The three black vertical dashed lines represent the 16th, 50th, and 84th percentiles. The blue lines and markers show the maximum likelihood values.}
    \label{fig:hor2corn}
\end{figure*}

\begin{figure*}
     \centering
    \includegraphics[width=0.75\textwidth]{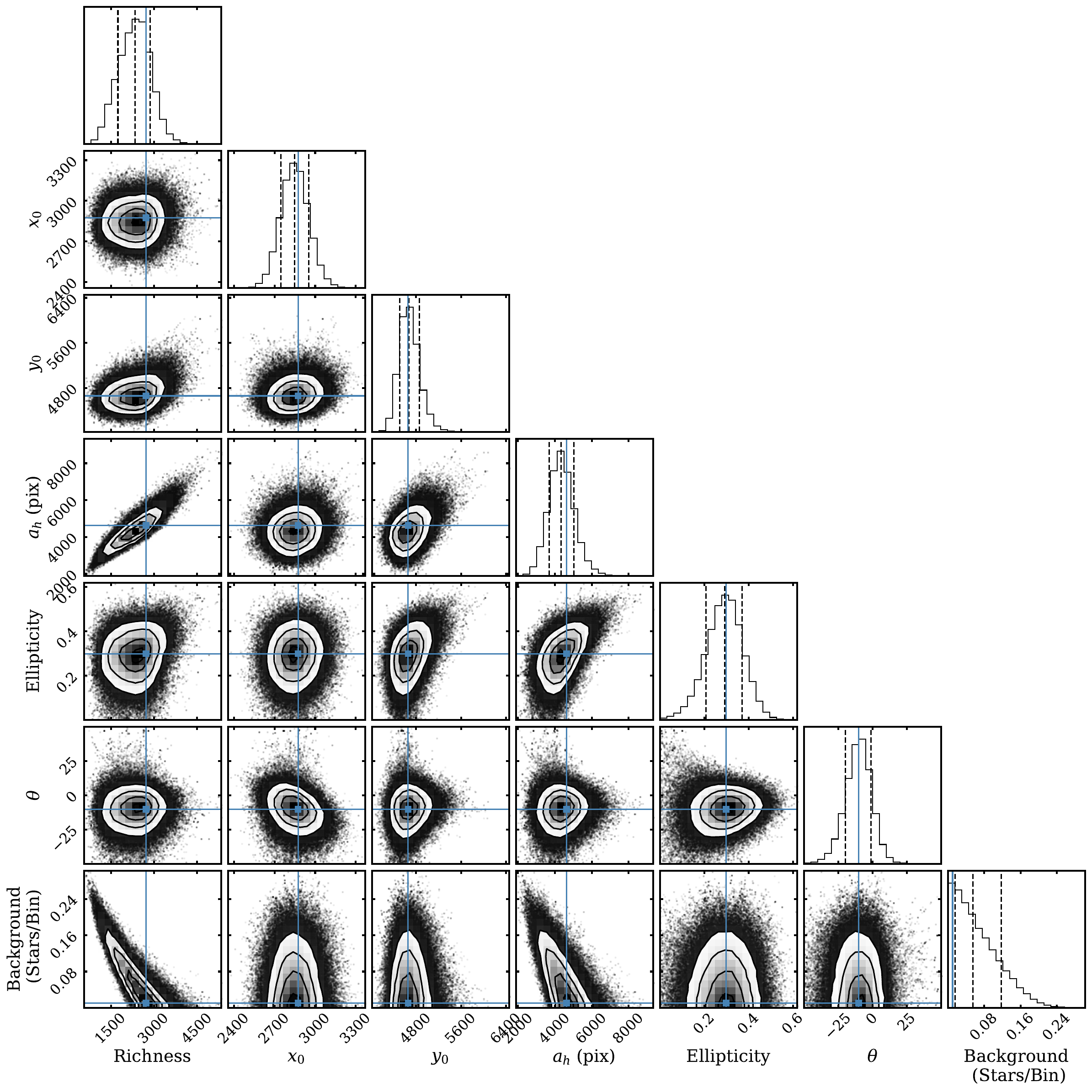}
    \caption{The posterior distributions of the 7-parameter structural fit for Hya~II using the exponential model. The three black vertical dashed lines represent the 16th, 50th, and 84th percentiles. The blue lines and markers show the maximum likelihood values.}
    \label{fig:hya2corn}
\end{figure*}

\begin{figure*}
     \centering
    \includegraphics[width=0.75\textwidth]{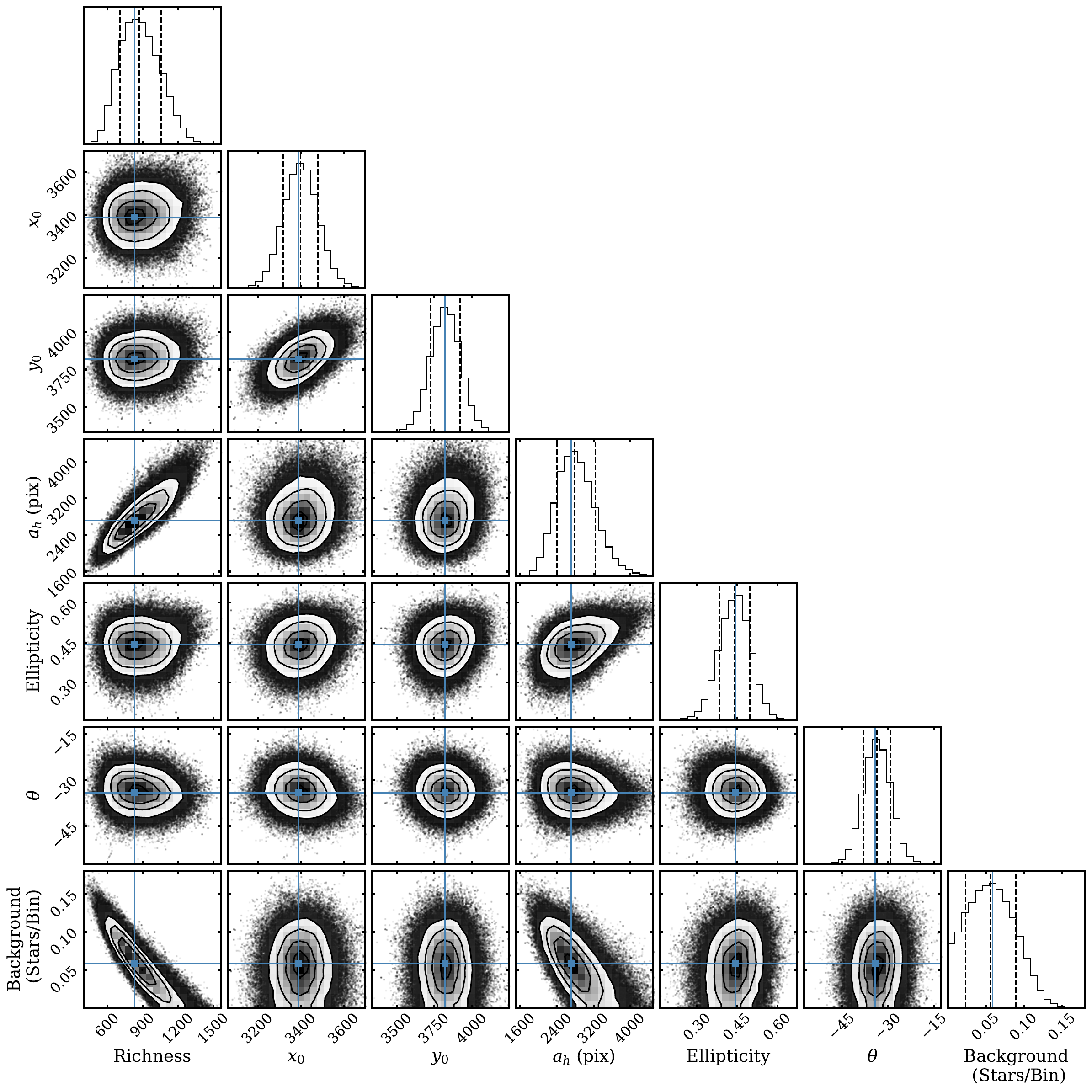}
    \caption{The posterior distributions of the 7-parameter structural fit for Phe~II using the exponential model. The three black vertical dashed lines represent the 16th, 50th, and 84th percentiles. The blue lines and markers show the maximum likelihood values.}
    \label{fig:phe2corn}
\end{figure*}

\begin{figure*}
     \centering
    \includegraphics[width=0.75\textwidth]{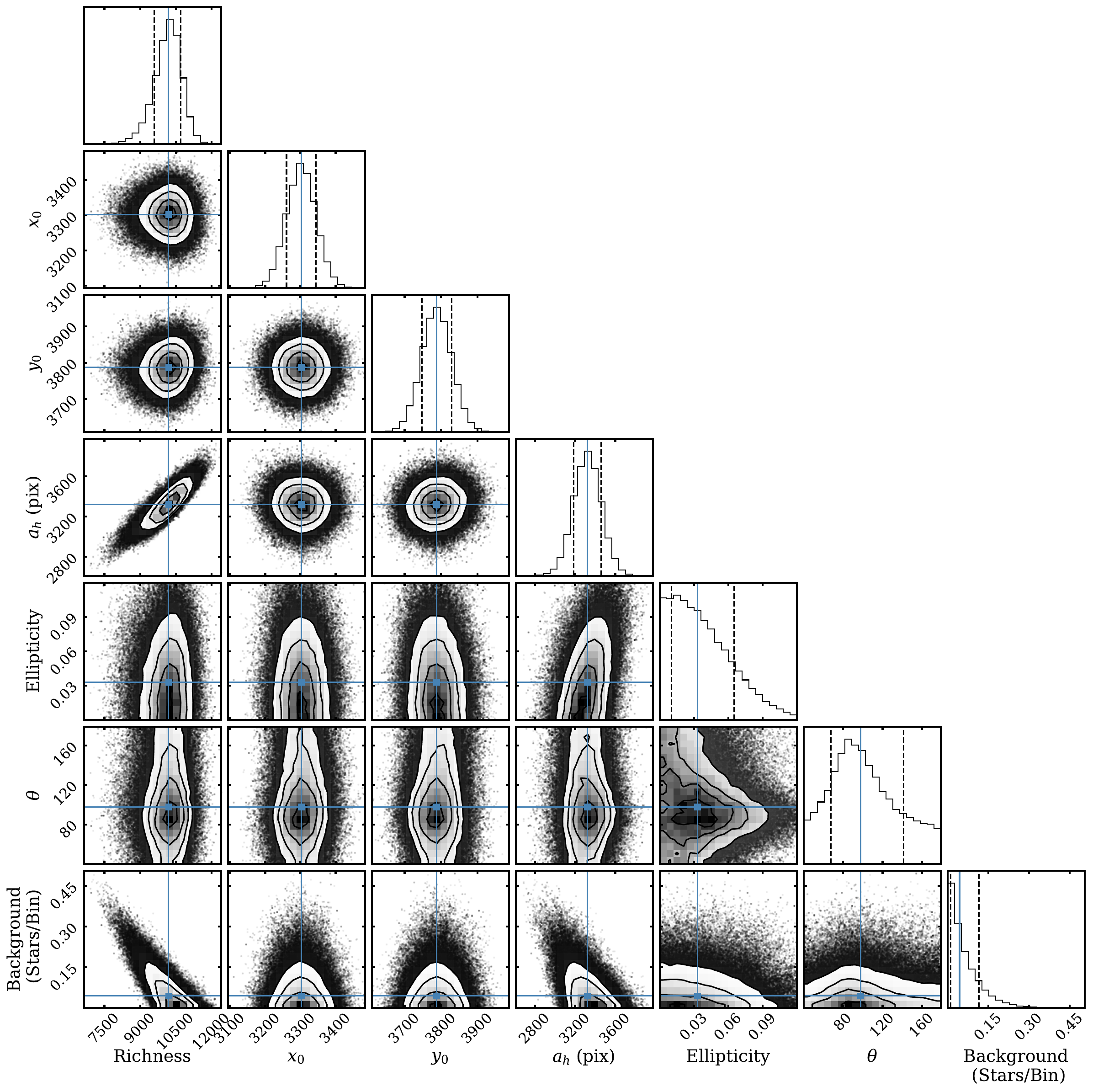}
    \caption{The posterior distributions of the 7-parameter structural fit for Sgr~II using the exponential model. The three black vertical dashed lines represent the 16th, 50th, and 84th percentiles. The blue lines and markers show the median values.}
    \label{fig:sgr2corn}
\end{figure*}

\begin{figure*}
     \centering
    \includegraphics[width=0.75\textwidth]{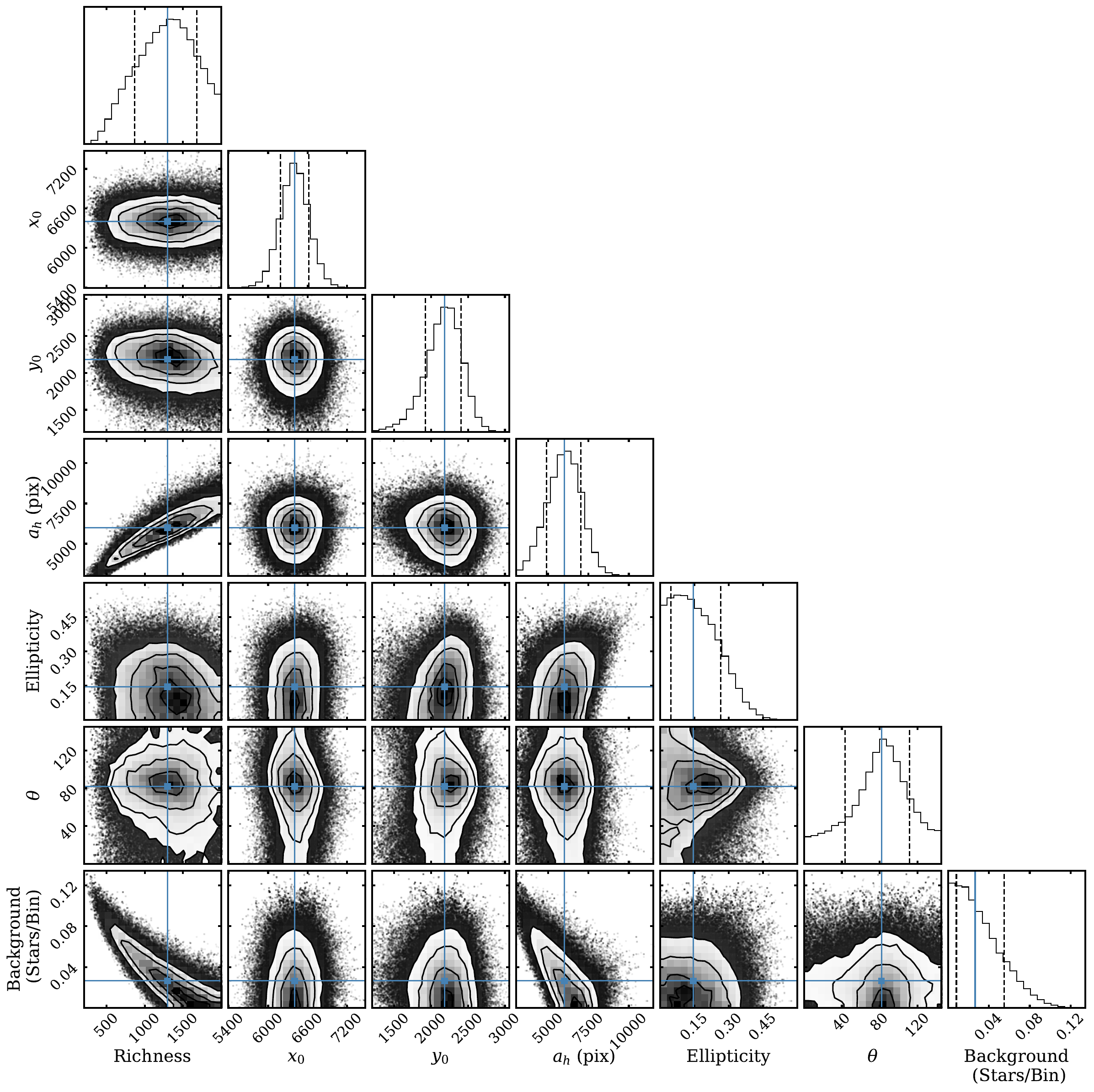}
    \caption{The posterior distributions of the 7-parameter structural fit for Tri~II using the exponential model. The three black vertical dashed lines represent the 16th, 50th, and 84th percentiles. The blue lines and markers show the median values.}
    \label{fig:tri2corn}
\end{figure*}

\section{References for Figure \ref{FIG:RHMV} } \label{app:ref}
We largely pulled from the compilations of \cite{simon2019} and \cite{mcConnachie2012} (2021 edition) for the UFD and UFD candidate parameters, but we have also included the more recently discovered satellites. Here we list the individual references.
\cite{majewski2003};
\cite{deJong2008}; 
\cite{drlica2015};
\cite{koposov2015b};
\cite{drlica2016};
\cite{crnojevic2016};
\cite{homma2016};
\cite{torrealba2016};
\cite{torrealba2016aq};
\cite{carlin2017};
\cite{luque2017};
\cite{homma2018};
\cite{koposov2018};
\cite{munoz2018};
\cite{mutluPakdil2018};
\cite{torrealba2018};
\cite{simon2020};
\cite{cerny2021a};
\cite{cerny2021b};
\cite{ji2021};
\cite{cerny2022};
\cite{richstein2022};
\cite{cerny2023};
\cite{smith2023}.

The globular cluster parameters come from the compilation \cite{harris2010}, with newer additions from \cite{munoz2018} and \cite{torrealba2019}.
The \cite{harris2010} compilation draws from the works of the following:
\cite{harris1984};
\cite{webbink1985};
\cite{peterson1987};
\cite{vbd1991};
\cite{tucholke1992};
\cite{mo1993};
\cite{trager1993};
\cite{cote1995}; 
\cite{trager1995};
\cite{harris1997};
\cite{kaisler1997};
\cite{lehmann1997};
\cite{ivanov2005};
\cite{kobulnicky2005};
\cite{mcLaughlin2005};
\cite{ferraro2006};
\cite{hilker2006};
\cite{mcLaughlin2006};
\cite{ortolani2006};
\cite{bellazzini2007}; 
\cite{carraro2007}; 
\cite{deMarchi2007};
\cite{froebrich2007};
\cite{koposov2007};
\cite{lan2007c};
\cite{lan2007a};
\cite{lan2007b};
\cite{bonatto2008}; 
\cite{kurtev2008};
\cite{carraro2009};
\cite{lan2010}.

We have also included the newer DELVE clusters from the following: 
\cite{mau2020};
\cite{cerny2022};
\cite{cerny2023delve}, and the faintest known MW satellite recently found by \cite{smith2024}.

\bibliography{main}{}
\bibliographystyle{aasjournal}



\end{document}

%% file: obsTable.tex
\begin{table*}
\caption{\\Summary of Primary ACS Observations and Field Completeness}
\begin{center}
\begin{tabular}{llllll}
\toprule
\multicolumn{1}{l}{Satellite Name} & UT Date & Filter & Exposure Time (s) & 50\% (VegaMag) & 90\% (VegaMag) \\
\midrule
 Grus~II (Gru~II) & 2016 Nov 02 & F606W & 4 $\times$ 1130 & 27.8 & 27.0 \\
 & 2016 Oct 08 & F814W & 4 $\times$ 1130 & 26.8 & 26.0 \\
 Horologium~I (Hor~I) & 2017 Jul 31 & F606W & 4 $\times$ 1131 & 27.8 & 27.0  \\
   & 2017 Aug 01 & F814W & 4 $\times$ 1131 & 26.8 & 26.0\\
Horologium~II East (Hor~II)  & 2016 Sep 26 & F606W &4 $\times$ 1131 & 28.1 & 27.3\\
 & 2016 Sep 26 & F814W &4 $\times$ 1131 & 27.1 & 26.3\\
Horologium~II West (Hor~II) & 2017 Mar 22 & F606W &4 $\times$ 1131 & 28.0 & 27.2 \\
 & 2017 Mar 22 & F814W &4 $\times$ 1131 & 27.0 & 26.2\\
 Hydra~II (Hya~II) & 2016 Dec 13 & F606W &4 $\times$ 1186 & 27.6 & 26.9\\
  & 2016 Dec 16 & F814W &4 $\times$ 1186 & 26.6 & 25.9\\
 Phoenix~II (Phe~II)  & 2017 May 08 & F606W &4 $\times$ 1131 & 27.8 & 27.0\\
    & 2017 May 08 & F814W &4 $\times$ 1131 & 26.8 & 26.0\\
 Reticulum~II (Ret~II) & 2017 Nov 18 & F606W &4 $\times$ 1145 & 27.4 & 26.2 \\
      & 2017 Nov 28 & F814W &4 $\times$ 1145 &26.4 &25.2\\
 Sagittarius~II (Sgr~II) & 2016 Oct 19 & F606W &4 $\times$ 1127 &27.5 &26.8\\
    & 2016 Oct 01 & F814W &4 $\times$ 1127 &26.8 &26.1\\
 Triangulum~II East (Tri~II) & 2016 Dec 31 & F606W &4 $\times$ 1137 &27.7 &27.0\\
& 2016 Dec 03 & F814W &4 $\times$ 1137 &26.7 &26.0\\
Triangulum~II West (Tri~II) & 2017 Jun 22 & F606W &4 $\times$ 1137 &27.7 &26.9\\
& 2017 Jun 22 & F814W &4 $\times$ 1137 &26.7 &25.9\\
Tucana~II Northeast (Tuc~II) & 2017 Jun 27 & F606W &4 $\times$ 1138 &28.1 &27.3\\
& 2017 Jun 29 & F814W &4 $\times$ 1138 &27.1 &26.3\\
Tucana~II Northwest (Tuc~II) & 2017 Aug 26 & F606W &4 $\times$ 1138 &27.9 &27.1\\
& 2017 Aug 31 & F814W &4 $\times$ 1138 &26.9 &26.1\\
Tucana~II Southeast (Tuc~II) & 2017 Aug 21 & F606W &4 $\times$ 1138 &28.0 &27.2\\
& 2017 Aug 25 & F814W &4 $\times$ 1138 &27.0 &26.2\\
Tucana~II Southwest (Tuc~II) & 2018 Mar 01 & F606W &4 $\times$ 1138 &27.5 &26.7\\
& 2018 Mar 03 & F814W &4 $\times$ 1138 &26.5 &25.7\\
Tucana~IV North (Tuc~IV) & 2017 Apr 06 & F606W &4 $\times$ 1140 &28.0 &27.3 \\
& 2017 Apr 06 & F814W &4 $\times$ 1140 &27.0 &26.3 \\
Tucana~IV South (Tuc~IV) & 2016 Sep 28 & F606W &4 $\times$ 1140 &27.9 &27.2 \\
 & 2016 Sep 28 & F814W &4 $\times$ 1140 &26.9 &26.2 \\
\bottomrule
\end{tabular}
\end{center}
\label{t:obs}
\end{table*}

%% file: derivedTable.tex
\begin{table*}
\centering
\begin{tabular}{llllllllll}
\multicolumn{10}{c}{\textbf{Adopted and Derived Parameters}}\\ 
\toprule
Parameter   & $(m-M)_0$$^a$  & Distance & $\sigma_v$$^{b,c}$  & $r_h$  & $r_h$ & $m_V$   & $M_{1/2}$ & ($M/L_V$)$_{1/2}$ & Ref.\\ 
Units & (mag) &(kpc) & (km s$^{-1}$) & (arcmin) & (pc) & (mag) & ($10^6$ $M_{\odot}$) & ($M_{\odot}/L_{\odot}$)\\\midrule
Hor~I & 19.5 $\pm0.2$ & 79$^{+8}_{-7}$ & 4.9$^{+2.8}_{-0.9}$ & 1.46 $\pm0.14$ & 34$^{+5}_{-4}$ & 16.12$^{+0.06}_{-0.09}$ & 0.88$^{+1.2}_{-0.35}$ & 890$^{+1200}_{-380}$ & 1,2 \\
Hor~II & 19.46 $\pm0.2$ & 78$^{+8}_{-7}$ & \multicolumn{1}{l}{...} & 1.43 $\pm0.17$ & 32 $\pm5$ & 17.39 $\pm0.10$ & \multicolumn{1}{l}{...} & \multicolumn{1}{l}{...} & 3\\
Hya~II & 20.64 $\pm0.16$ & 134$\pm10$ & $<$3.6,$<$4.5 & 2.26$^{+0.37}_{-0.34}$ & 88$^{+16}_{-14}$ & 15.80$^{+0.11}_{-0.22}$ & $<$2.0  & $<$670  & 4,5\\
Phe~II & 19.85$^{+0.16}_{-0.17}$ & 93$\pm7$ & 11.0 $\pm9.4$  & 1.18$^{+0.20}_{-0.18}$ & 32$^{+6}_{-5}$ & 17.00$^{+0.13}_{-0.09}$ & 4.8$^{+11.6}_{-3.7}$ & 7800$^{+18000}_{-6500}$ & 6,7  \\
Sgr~II & 19.2 $\pm0.2$ & 69$^{+7}_{-6}$ & 1.7 $\pm0.5$ & 1.82$^{+0.08}_{-0.09}$ &  37 $\pm4$ & 13.82$^{+0.06}_{-0.04}$ & 0.10$^{+0.07}_{-0.05}$ & 17$^{+12}_{-8}$ & 8,9  \\
Tri~II & 17.27 $\pm0.11$  & 28$^{+2}_{-1}$ & $<$3.4,$<$4.2 & 3.34$^{+0.60}_{-0.61}$ & 28 $\pm5$ & 15.97$^{+0.23}_{-0.25}$ & $<$0.48  & $<$3200  & 10,11       \\ \midrule
Gru~II & 18.71 $\pm0.1$ & 55$\pm3$ & $<$1.9,$<$2.0 & 5.24$^{+0.50}_{-0.48}$ & 84$\pm9$ & \multicolumn{1}{l}{...} & $<$0.35 & $<$330 & 12,13 \\
Ret~II & 17.5 $\pm0.1$ & 32$^{+2}_{-1}$ & 2.97$^{+0.43}_{-0.35}$ & 3.98$^{+0.53}_{-0.57}$ & 37 $\pm5$ & \multicolumn{1}{l}{...} & 0.48$^{+1.5}_{-1.1}$ & 640$^{+200}_{-150}$ & 8,14\\
Tuc~II & 18.8 $\pm0.2$ & 58$^{+6}_{-5}$ & 3.8$^{+1.1}_{-0.7}$ & 9.83$^{+1.66}_{-1.11}$ & 170$^{+37}_{-33}$  & \multicolumn{1}{l}{...} & 2.9$^{+2.0}_{-1.1}$ & 2100$^{+1400}_{-780}$ & 1,15\\
Tuc~IV & 18.36$^{+0.18}_{-0.19}$ & 47$\pm4$ & 4.3$^{+1.7}_{-1.0}$ & 7.26$^{+1.14}_{-0.88}$ & 100$^{+17}_{-15}$ &  \multicolumn{1}{l}{...} & 2.2$^{+1.8}_{-1.1}$ & 3100$^{+2900}_{-1600}$ & 13\\
\bottomrule
\end{tabular}
\begin{FlushLeft}
\caption{For each target, we list the distance modulus and velocity dispersion values from the literature. For the first six targets, we also list the azimuthally-averaged $r_h$ in arcminutes and parsecs, as well as our derived apparent and absolute $V$-band magnitudes. We list the dynamical mass within the 2D projected $a_h$ ($M_{1/2}$) and mass-to-light ratio (($M/L_V$)$_{1/2}$) derived using our $r_h$ and $M_V$ values. For the last four rows, we list the literature properties for the targets that we were unable to measure.}
\label{tab:derived}
\footnotesize{$^a$\cite{koposov2015b} state that the uncertainties on their distance moduli are 0.1-0.2. We adopt 0.2 in our analysis. \\
$^b$\cite{kirby2015,kirby2017} report their 90\% and 95\% confidence levels for the velocity dispersions.} \\
$^c$ \cite{simon2020} report the 90\% and 95.5\% confidence upper limits on the velocity dispersion.\\
\footnotesize{$^1$\cite{koposov2015b}; $^2$\cite{koposov2015a}; 
$^3$\cite{kim2015};
$^4$\cite{martin2015}; $^5$\cite{kirby2015}; 
$^6$\cite{nagarajan2022}; $^7$\cite{fritz2019}; $^8$\cite{mutluPakdil2018}; $^9$\cite{longeard2021}; 
$^{10}$\cite{carlin2017}; $^{11}$\cite{kirby2017}; $^{12}$\cite{martinezVel2019}; 
$^{13}$\cite{simon2020};
$^{14}$\cite{ji2023};$^{15}$\cite{chiti2023}
}
\end{FlushLeft}
\end{table*}

%% file: simTable.tex
\begin{table*}
\centering
\begin{tabular}{lllllll}
\multicolumn{7}{c}{\textbf{Simulation Descriptions}}\\
\toprule
Simulation Set & $N_{\rm gal}$ $^a$ & Hydro. Code & Baryonic Res. &  Halo Mass$^c$, $z=0$ &Environment & Ref.\\
 & & & ($M_{\odot}$) & ($M_{\odot}$) & &\\ \hline
\multicolumn{7}{c}{\textbf{CDM}}\\ 
DC Justice League & 36 & CHANGA & 994 & $1.1\times10^6 - 3.7\times10^9$ &MW-like & 1\\ 
EDGE & 31 & RAMSES-RT & $20-300$ & $8.0\times10^8 - 3.7\times10^9$  &Isolated & 2, 3, 4, 5\\ 
FIRE-2 + CDM & 42 & GIZMO & $30-250$ & $7.0\times10^8 - 9.0\times10^9$ &Isolated & 6, 7, 8, 9\\
FIRE-2, Jahn 2022 & 25 & GIZMO & 880 & $(2.9 - 6.1)\times10^8$ &LMC-like & 6, 10, 11\\
GEAR & 19 & GEAR$^b$ & 380 & $3.0\times10^8 - 2.4\times10^9$ &Isolated & 12\\
Jeon 2017 & 6 & Modified GADGET & 465 & $(1.5-4.0)\times10^9$ &Field & 13 \\
Jeon 2021 & 42 & Modified GADGET & 65 & $(2.0-2.5)\times10^8$ &Field & 14, 15\\
LYRA & 4 & AREPO & 4 & $3.0\times10^8-2.8\times10^9$ &Isolated & 16\\
Macci\`o 2017, Cent. & 9 & GASOLINE & $40, 135$ & $(1.7-5.7)\times10^9$  &Isolated overdensity & 17\\
Macci\`o 2017, Sat. & 6 & GASOLINE & $90, 305$ & $4.5\times10^8-1.1\times10^{10}$ &Isolated overdensity & 17\\
Macci\`o 2019 (CDM) & 4 & GASOLINE-2 & $2.4\times10^3-1.9\times10^4$ & $(2.9-9.1)\times10^9$ &Isolated overdensity & 18 \\
Marvel-ous Dwarfs & 32 & CHANGA & 420 & $2.6\times10^8 - 4.5\times10^9$ &Local Volume &  19\\ 
TNG (CDM) & 47 & AREPO & $4.2\times10^4$ & $2.2\times10^7 - 2.8\times10^9$ &Field & 20\\\hline
\multicolumn{7}{c}{\textbf{SIDM}}\\ 
FIRE-2 + SIDM & 4 & GIZMO & 500 & $8.1\times10^9$ &Isolated & 21\\
TNG (ETHOS/SIDM) & 52 & AREPO &  $4.2\times10^4$ & $1.9\times10^6 - 4.2\times10^9$ &Field & 22\\ \hline
\multicolumn{7}{c}{\textbf{WDM}}\\ 
FIRE-2 + WDM & 3 & GIZMO & 500 & $(6.7-8.5)\times10^9$ &Isolated &  23\\
Macci\`o 2019 (WDM) & 3 & GASOLINE-2 & $2.4\times10^3-1.9\times10^4$ & $3.1\times10^9 - 1.3\times10^{10}$ &Isolated overdensity & 18\\
\bottomrule
\end{tabular}
\begin{FlushLeft}
\caption{For each set of simulations shown in Figure \ref{fig:rhMV}, we list the set name, number of galaxies shown in the plot, the hydrodynamics code used, the baryonic resolution, the $z=0$ DM halo mass range, the environment in which the galaxies were simulated, and the references.}
\footnotesize{
$^a$The number of galaxies shown in Figure \ref{fig:rhMV}, with an $M_V\gtrsim-10$ cut imposed.\\
$^b$GEAR was based on GADGET-2 \citep{springel2005}. \\
$^c$DM halo mass range (rounded to two significant figures) for the galaxies shown in \ref{fig:rhMV}. Note that the mass range for the full simulation set could be larger.\\
$^1$\cite{Applebaum2021}; 
$^2$\cite{agertz2020}; 
$^3$\cite{rey2020};
$^4$\cite{orkney2021};
$^5$\cite{prgomet2022};
$^6$\cite{fitts2017};
$^7$\cite{hopkins2018};
$^8$\cite{wheeler2019};
$^9$\cite{wetzel2023};
$^{10}$\cite{elBadry2018};
$^{11}$\cite{jahn2022};
$^{12}$\cite{sanati2023};
$^{13}$\cite{Jeon2017};
$^{14}$\cite{Jeon2021a};
$^{15}$\cite{Jeon2021b};
$^{16}$\cite{gutcke2022};
$^{17}$\cite{maccio2017};
$^{18}$\cite{maccio2019};
$^{19}$\cite{munshi2021};
$^{20}$\cite{rose2023};
$^{21}$\cite{robles2017};
$^{22}$J.~Rose et al.~2024, in prep.;
$^{23}$\cite{bozek2019}
}
\label{tab:simtable}
\end{FlushLeft}
\end{table*}